\documentclass{article}

\usepackage{amssymb, amsmath, mathrsfs, amsfonts, graphicx, subfigure, natbib}
\usepackage{adjustbox}
\usepackage{array, epsfig, rotating, pdflscape, bm}
\usepackage{algorithm, algpseudocode}
\usepackage{enumerate}
\usepackage{etoolbox}

\numberwithin{equation}{section}

\newcommand{\be} {\mathbf {e}}
\newcommand{\bD} {\mathbf {D}}
\newcommand{\bI} {\mathbf {I}}

\newcommand{\bP} {\mathbf {P}}

\newcommand{\br} {\mathbf {r}}

\newcommand{\bu} {\mathbf {u}}
\newcommand{\bU} {\mathbf {U}}
\newcommand{\bW} {\mathbf {W}}
\newcommand{\bV} {\mathbf {V}}
\newcommand{\bv} {\mathbf {v}}
\newcommand{\bX} {\mathbf {X}}

\newcommand{\bY} {\mathbf {Y}}
\newcommand{\by} {\mathbf {y}}
\newcommand{\bZ} {\mathbf {Z}}
\newcommand{\bmu} {\mbox{\boldmath $\mu$}}
\newcommand{\bgamma} {\mbox{\boldmath $\gamma$}}

\newcommand{\bOmega} {\mbox{\boldmath $\Omega$}}
\newcommand{\bphi} {\mbox{\boldmath $\phi$}}

\newcommand{\bone} {\mathbf {1}}
\newcommand{\bzero} {\mathbf {0}}
\def\T{{ \mathrm{\scriptscriptstyle T} }}

\providecommand{\keywords}[1]
{
  \small	
  \textbf{\textit{Keywords---}} #1
}

\title{Elliptically symmetric distributions for directional data of arbitrary dimension}
\author{Zehao Yu and Xianzheng Huang\\ Department of Statistics, University of South Carolina \\
e-mail: zehaoy@email.sc.edu; huang@stat.sc.edu}
\date{}

\begin{document}

\maketitle

\begin{abstract}
We formulate a class of angular Gaussian distributions that allows different degrees of isotropy for directional random variables of arbitrary dimension. Through a series of novel reparameterization, this distribution family is indexed by parameters with meaningful statistical interpretations that can range over the entire real space of an adequate dimension. The new parameterization greatly simplifies maximum likelihood estimation of all model parameters, which in turn leads to theoretically sound and numerically stable inference procedures to infer key features of the distribution. Byproducts from the likelihood-based inference are used to develop graphical and numerical diagnostic tools for assessing goodness of fit of this distribution in a data application. Simulation study and application to data from a hydrogeology study are used to demonstrate implementation and performance of the inference procedures and diagnostics methods.
\end{abstract}

\keywords{Angular Gaussian distribution, compositional data, maximum likelihood, residual}

\section{Introduction}
\label{sec:intro}

Directional data are ubiquitous in oceanography with wave directions as an example, in meteorology where wind directions are directional data of interest, and in biology where protein backbone structures are directional data researchers study. These exemplify directional data of dimension no higher than three. Other examples of low dimensional direction data include migratory movements of animals, and measurements on a periodic scale, such as weekdays and hours. Directional data of higher dimensions arise in bioinformatics and hydrogeology, among many other fields of research. For example, gene expression data associated with a large number of genes for each experimental unit are often standardized to preserve directional characteristics when studying the fluctuation of gene expressions over cell cycles \citep{dortet2008model}. By transforming the original gene expression data on a high dimensional Euclidean space to a unit hypersphere, one ignores absolute expression levels and can obtain better clustering of genes that are functionally related \citep{banerjee2005clustering}. Another example of directional data with dimension usually higher than three is compositional data \citep{pawlowsky2006compositional, alenazi2021review}. For instance, microbiome data are often summarized as the composition of bacterial taxa so that one can focus on the microbial relative abundances as opposed to absolute abundances in microbiome analysis \citep{shi2016regression}. A compositional data point is a vector with non-negative components that sum to one, hence a component-wise square-root transformation of this vector yields a vector on a unit hypersphere \citep{scealy2011regression}.

Each of the above examples of directional data can be viewed as realizations of a random variable supported on a unit-radius  $d$-dimensional spherical space defined by $\mathbb{S}^{d-1} = \lbrace \by\in \mathbb{R}^d : \|\by\| = 1 \rbrace$, for $d\geq 2$, where $\|\by\|$ is the $L_2$-norm of $\by$. \citet{lee2010circular} provided a brief survey of statistical methods for analyzing circular data, i.e., directional data with $d=2$. Two general strategies for constructing a circular distribution are highlighted in this review paper: one uses a ``wrapped" circular version of a random variable supported on $\mathbb{R}$ to formulate a circular distribution; the other deduces a circular distribution via projecting a univariate random variable on $\mathbb{R}$ or a bivariate random variable on $\mathbb{R}^2$ onto the circle. Both strategies have been generalized and used to formulate directional distributions on $\mathbb{S}^{d-1}$ for $d>2$. With the Gaussian distribution playing an important role in statistics, it is not surprising that directional distributions originating from a Gaussian distribution have been most studied and adopted in practice, including the so-called wrapped normal distribution and projected normal distribution, with more attention on the latter in recent literature. In particular,   \citet{presnell1998projected} used a projected multivariate normal distribution to construct a regression model for a circular response and linear predictors, and employed the maximum likelihood method to infer unknown parameters. \citet{wang2013directional} incorporated projected normal distributions to develop Bayesian
hierarchical models for analyzing circular data. 
\citet{hernandez2017general} proposed Bayesian inferential method for directional data of arbitrary dimension, again modelled by projected normal distributions.

Projected normal distributions are also referred to as angular Gaussian distributions. Different angular Gaussian distributions are created by imposing different constraints on the parameter space associated with a multivariate Gaussian distribution in order to resolve the non-identifiability issue that arises when the support of a random variable changes from a Euclidean space to a spherical space. \citet{paine2018elliptically} imposed constraints on the mean vector and variance-covariance matrix of a Gaussian distribution so that the resultant angular Gaussian distribution is identifiable and, more interestingly,  elliptically symmetric. The authors thus coined their proposed distribution as the elliptically symmetric angular Gaussian distribution, ESAG for short. \citet{paine2020spherical} further developed regression models for directional data assuming an ESAG distribution for the response given covariates. Both works on ESAG focus on directional data with $d\le 3$. More recently,  \citet{scealy2019scaled} proposed a new directional distribution, called scaled von Mises-Fisher distribution, using grouped transformations of the von Mises-Fisher distribution to achieve elliptical symmetry. The authors used this new distribution to model archeomagnetic data that can be converted to directional data with $d=3$. The feature of elliptical symmetry of a distribution makes capturing certain anisotropic pattern of directional data possible. An added benefit of ESAG is that the normalization constant in its probability density function is much easier to compute compared to many existing directional distributions, such as the Kent distribution \citep{kent1982fisher}. This makes maximum likelihood estimation under the ESAG model for directional data more straightforward. 

To incorporate the constraints imposed on the mean vector and variance-covariance matrix of a Gaussian distribution when formulating ESAG, \citet{paine2018elliptically} designed a parameterization of ESAG when $d=3$, which allows one to bypass the complicated problem of optimization with constraints when finding the maximum likelihood  estimators of the induced  parameters. But their parameterization cannot be easily generalized to cases with $d>3$. This limits the use of ESAG in applications where directional data of higher dimension are observed. The first contribution of our study presented in this paper is a novel parameterization of ESAG that yields a mathematically sophisticated  model for directional data of arbitrary dimension. This new parameterization of ESAG for $d\ge 3$ is presented in Section~\ref{sec:parameterization}. 
Under the new parameterization, maximum likelihood estimation translates to a routine numerical problem of optimization without constraints, as we describe in Section~\ref{sec:mle}. A legitimate concern in any parametric modelling is potential violations of certain model assumptions in a given application. To address this concern, we propose model diagnostics methods that exploit directional residuals in Section~\ref{sec:verify}, which constitutes a second major contribution of our study. Operating characteristics of the proposed model diagnostics methods are demonstrated in simulation study in Section~\ref{sec:sim}. In Section~\ref{sec:realdata}, we entertain data  from hydrogeological research, where we fit ESAG to compositional  data from different geographic locations. Section~\ref{sec:conc} summarizes the contributions of the study and outlines the follow-up research agenda.

\section{The ESAG distribution}
\label{sec:parameterization}
\subsection{Constraints on parameters}
\label{sec:constraints}
Let $\bX$ be a $d$-dimensional Gaussian variable with mean $\bmu$ and variance-covariance $\bV$, i.e., $\bX\sim N_d (\bmu, \bV)$. Then the normalized variable, $\bY=\bX/\|\bX\|$, follows an angular Gaussian distribution, $\mbox{AG}(\bmu, \bV)$, supported on $\mathbb{S}^{d-1}$. Parameters in $\bmu$ and $\bV$ associated with $\mbox{AG}(\bmu, \bV)$ are not identifiable because $\bX/\|\bX\|$ and $c\bX/\|c\bX\|$ are equal for $c>0$, and thus they follow the same angular distribution, even though $\bX$ and $c\bX$ have different mean or/and variance-covariace when $c\ne 1$. To construct an identifiable angular Gaussian distribution,  \citet{paine2018elliptically} impose the following two sets of constraints on $\bmu$ and $\bV$, where $\mbox{det}(\cdot)$ refers to the determinant of a matrix, 
\begin{align}
    \bV\bmu & = \bmu, \label{eq:constraint1}\\ \mbox{det}(\bV) & =1, 
\label{eq:constraint2}
\end{align}
leading to the ESAG distribution, with the probability density function given by \begin{equation}
f(\by|\bmu,\bV)=\frac{(2\pi)^{-(d-1)/2}}{ (\by^\T \bV^{-1} \by)^{d/2}}\exp \left[ \frac{1}{2}\left\{\frac{(\by^\T\bmu)^2}{\by^\T \bV^{-1}\by} -\bmu^\T \bmu\right\}\right] M_{d-1}\left\{\frac{\by^\T\bmu}{(\by^\T\bV^{-1}\by)^{1/2}}\right\},
\label{eq:pdf}
\end{equation}
where $M_{d-1}(t)=(2\pi)^{-1/2}\int_{0}^{\infty}x^{d-1} \exp\{ -(x-t)^2/2 \} dx$. Henceforth, we say that $\bY$ follows a $(d-1)$-dimensional ESAG, or $\bY\sim \mbox{ESAG}_{d-1}(\bmu, \bV)$, if $\bY$ follows a distribution specified by the density in (\ref{eq:pdf}) with constraints in (\ref{eq:constraint1}) and (\ref{eq:constraint2}). 

\begin{figure}[h]
\centering
\subfigure[]{\includegraphics[width=1.5in]{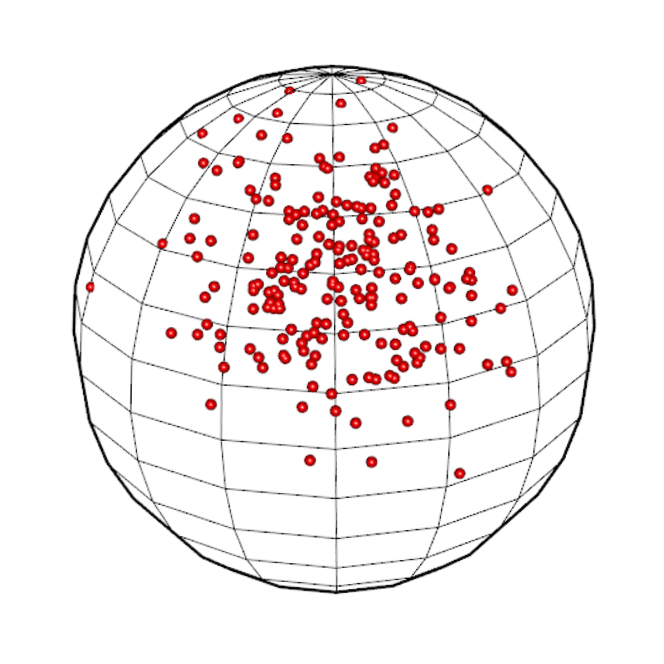}}
\subfigure[]{\includegraphics[width=1.5in]{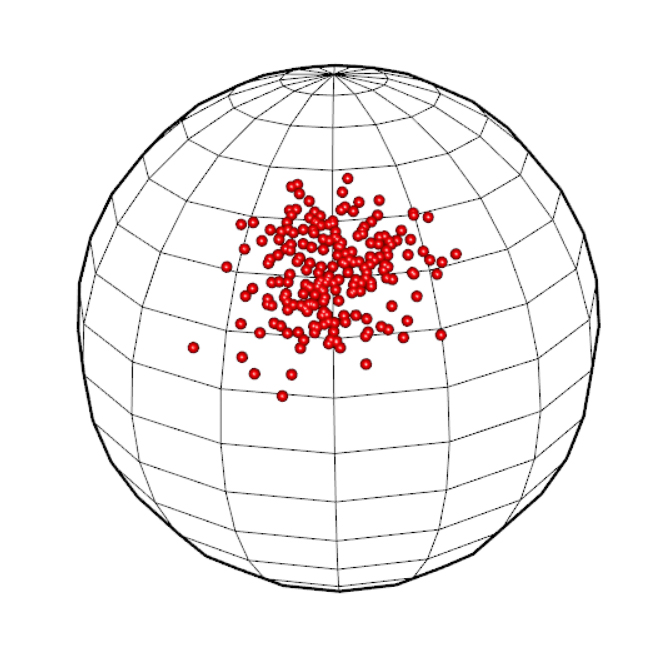}}\\
\subfigure[]{\includegraphics[width=1.5in]{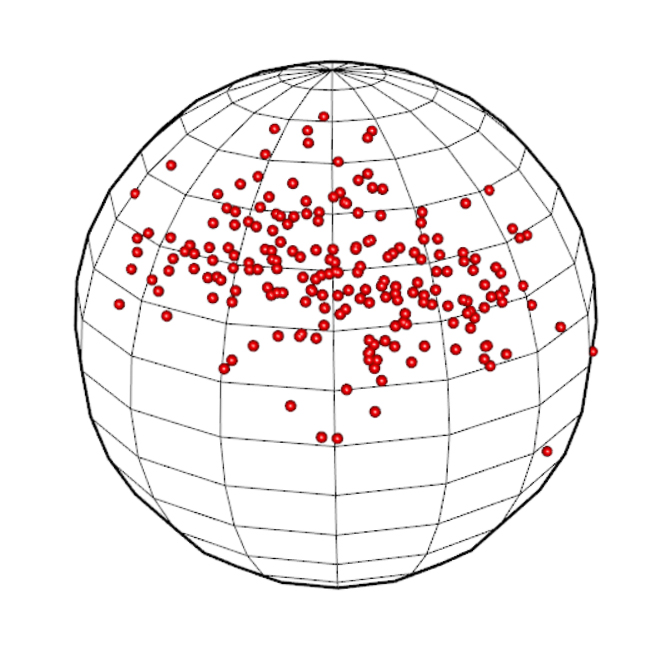}}
\subfigure[]{\includegraphics[width=1.5in]{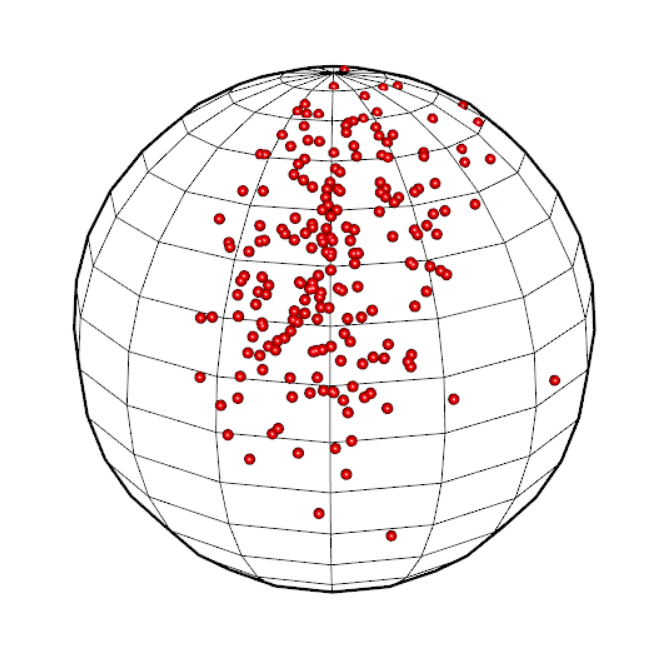}}
\caption{\label{fig:ESAGdata}Four random samples from $\mbox{ESAG}_2(\bmu, \, \bV)$ with $\bmu$ and $\bV$ specified by (a)--(d) in Section~\ref{sec:constraints}.}
\end{figure}

Figure~\ref{fig:ESAGdata} presents four random samples scattering on 3-dimensional spheres, generated from $\mbox{ESAG}_2(\bmu, \, \bV)$ with the following parameters specifications, where $\bone_d$ is a vector of $d$ ones and $\bI_d$ is $d$-dimensional identity matrix: 
\vspace{0.5cm}

\noindent\begin{minipage}{.5\linewidth}
\begin{itemize}
    \item[(a)]$\bmu\ =\ 2\times \bone_3, \, \ \ \bV\  =\ \bI_3$; 
    \item[(c)]$\bmu\ =\ 2\times \bone_3, \\ \bV = \begin{bmatrix}
1.57 & -0.08 & -0.50 \\
-0.08 & 0.74 & 0.34 \\
-0.50 & 0.34 & 1.16 
\end{bmatrix}$;
\end{itemize}

\end{minipage}%
\begin{minipage}{.5\linewidth}
\begin{itemize}
    \item[(b)]$\bmu\ =\ 4\times \bone_3, \,\ \  \bV\ =\ \bI_3$; 
      \item[(d)]$\bmu\ =\ 2\times \bone_3, \\ \bV = 
\begin{bmatrix}
0.74 & -0.08 & 0.34 \\
-0.08 & 1.57 & -0.50 \\
0.34 & -0.50 & 1.16 
\end{bmatrix}$.
\end{itemize}
\end{minipage}
\vspace{0.5cm}

\noindent Comparing the four data clouds depicted in Figure~\ref{fig:ESAGdata}, one can see that a larger $\|\bmu\|$ leads to less variability in a random sample (e.g., contrasting (a) with (b)); and $\bV$ also influences the orientation of the data cloud (e.g., comparing (a), (c), and (d)).

Because the dimension of the parameter space associated with $N_d(\bmu, \bV)$ is $d(d+3)/2$, and there are $d+1$ constraints imposed by (\ref{eq:constraint1}) and (\ref{eq:constraint2}), there are at most $p=(d-1)(d+2)/2$ identifiable parameters for $\mbox{ESAG}_{d-1}(\bmu, \bV)$. Let $\bOmega$ be the $p\times 1$ parameter vector that specifies $\mbox{ESAG}_{d-1}(\bmu, \bV)$. To facilitate likelihood-based inference, it is desirable to formulate $\bOmega$ so that the parameter space is $\mathbb{R}^p$. For this purpose, we define $\bOmega=(\bmu^\T, \bgamma^\T)^\T$, where, clearly, $\bmu\in \mathbb{R}^d$, and thus $\bgamma\in \mathbb{R}^{(d-2)(d+1)/2}$ includes parameters needed to specify $\bV$ that satisfies (\ref{eq:constraint1}) and (\ref{eq:constraint2}) after $\bmu$ is given. 

The parameterization leading to $\bgamma$ starts from the spectral decomposition of $\bV$, 
\begin{align}
    \bV = \sum_{j=1}^{d} \lambda_j \bv_j\bv_j^\T, \label{eq:Vdecom}
\end{align}
where $\lambda_1,...,\lambda_d\in (0,\, +\infty)\triangleq\mathbb{R}_+$ are  eigenvalues of $\bV$, and $\bv_1,...,\bv_d$ are the corresponding orthonormal eigenvectors. According to (\ref{eq:constraint1}), one of the eigenvalues of $\bV$ is equal to 1, with $\bmu$ being the corresponding (non-zero) eigenvector. Without loss of generality, we set $\lambda_d = 1$ and $\bv_d = \bmu/\|\bmu\|$. It follows that   $\lambda_1 = 1/\prod_{j=2}^{d-1}\lambda_j$ since $\mbox{det}(\bV)=\prod_{j=1}^d \lambda_j=1$ by (\ref{eq:constraint2}). To this end, once $\bmu$ is given, one needs to formulate $\bgamma$ so that it can be mapped to $ \lambda_2,...,\lambda_{d-1}$ and  $\bv_1,...,\bv_{d-1}$, through which $\bV$ is determined via (\ref{eq:Vdecom}).
In what follows, we present the derivations leading to such mapping in two steps. 

\subsection{Constructing eigenvectors}
\label{sec:basis}
We first define an orthonormal basis of $\mathbb{R}^d$ as a function of $\bmu=(\mu_1, \ldots, \mu_d)^\T$, denoted by $(\tilde \bv_1, \ldots, \tilde \bv_d)$, with $\tilde \bv_j=\bu_j/\|\bu_j\|$, for $j=1, \ldots, d$, and 
\begin{equation}
\label{eq:vtilde}
\begin{aligned}
\bu_j & = 
\left\{
\begin{array}{ll}
(-\mu_2,\, \mu_1,\,0,...,0)^\T, & \mbox{for $j=1$,}\\
(\mu_1\mu_{j+1},\, ...,\, \mu_j\mu_{j+1},\,- \sum_{k=1}^{j} \mu_k^2, \, 0, \ldots, 0)^\T, & \mbox{for $j=2,...,d-1$,} \\
\bmu & \mbox{for $j=d$}.
\end{array}
\right.
\end{aligned}
\end{equation}
If (\ref{eq:vtilde}) yields $\bu_j=\bzero_d$, for $j\in \{1, \ldots, d-1\}$, then we set $\bu_j=\be_j$, i.e., the unit vector with 1 at the $j$-th entry.

By (\ref{eq:vtilde}), $\tilde \bv_d=\bv_d$. We next relate $\{\tilde \bv_j\}_{j=1}^{d-1}$ to $\{\bv_j\}_{j=1}^{d-1}$ via a $(d-1)$-dimensional rotation matrix $\mathcal{R}_{d-1}$ formulated following the strategy proposed by \citet{murnaghan1952element}, which  \citet{scealy2011regression} exploited to parameterize the Kent distribution for modeling compositional data. Following this strategy, for $d>3$, we write $\mathcal{R}_{d-1}$ as a product of $(d-2)(d-1)/2$ plane rotation matrices that are functions of longitude angles, $ \theta_1,..,\theta_{d-2}\in [-\pi , \pi )$, and latitude angles, $\phi_1,..,\phi_{(d-2)(d-3)/2}\in [0,\pi]$. Here, a $(d-1)$-dimensional plane rotation matrix $R_{jk}^*(\cdot)$ comes from replacing the $(j,j)$, $(j,k)$, $(k,j)$, and $(k,k)$ entries of $\bI_{d-1}$ by  $\cos(\cdot)$, $-\sin(\cdot)$, $\sin(\cdot)$, and $\cos(\cdot)$, respectively. More specifically, we define $(\bv_1,...,\bv_{d-1})=(\tilde{\bv}_1,...,\tilde{\bv}_{d-1})\mathcal{R}_{d-1}$, where 
\begin{equation}
\mathcal{R}_{d-1} =\left[\prod_{m=1}^{d-3}\left\{ R_{12}^*(\theta_{d-m-1})\prod_{j=1}^{d-m-2}R_{j+1,j+2}^*(\phi_{1-j+(d-m-1)(d-m-2)/2})  \right\}  \right]R_{12}^*(\theta_1).
\label{eq:rotation}
\end{equation}

The rotation matrix in (\ref{eq:rotation}) depends on  $(d-2)(d-1)/2$ angles that we refer to as orientation parameters in the sequel. Putting the orientation parameters along with the eigenvaules, we have the collection of parameters needed to specify $\bV$ after $\bmu$ is given in $(\lambda_2, \ldots, \lambda_{d-1}, \, \theta_1, \ldots, \theta_{d-2}, \, \phi_1, \ldots, \phi_{(d-2)(d-3)/2})$. We next turn to defining $\bgamma \in \mathbb{R}^{(d-2)(d+1)/2}$ that can be mapped to this collection of parameters via groups of transformations. 

\subsection{Grouped spherical transformations}
\label{sec:grouping}
Following setting $\lambda_d=1$, we now let $\lambda_1 \le ...\le \lambda_{d-1}$, and write $\lambda_j=(r_{j-1}+1)\lambda_{j-1}$, where $r_{j-1}\ge 0$, for $j=2, \ldots, d-1$. Because $\prod_{j=1}^{d-1} \lambda_j=1$ by (\ref{eq:constraint2}), the first $d-1$ eigenvalues can be expressed in terms of $r_1, \ldots, r_{d-2}$ as follows,  
\begin{equation}
\lambda_1 = \left\{ \prod_{j=1}^{d-2} (r_j+1)^{d-(j+1)}\right\}^{-1/(d-1)} \textrm{ and } 
\lambda_j = \lambda_1\prod_{k=1}^{j-1}(r_k+1),  \mbox{ for $j=2,...,d-1$.} 
\label{eq:lambdas}
\end{equation}
We call $r_1, \ldots, r_{d-2}$  radial parameters for a reason to become clear momentarily. In what follows, we define transformations mapping $\bgamma$ to radial and orientation parameters in $\widetilde\bOmega=(r_1, \ldots, r_{d-2}, \, \theta_1, \ldots, \theta_{d-2}, \, \phi_1, \ldots, \phi_{(d-2)(d-3)/2})^\T$ after partitioning these parameters into $d-2$ groups motivated by the following observations. 

As the dimension of $\bY$ increases from $k$ to $k+1$, where $k \ge 3$, we need one additional radial parameter to account for the additional eigenvalue of $\bV$, along with, by (\ref{eq:rotation}), one additional longitude angle and $k-2$ additional latitude angles, yielding a total of $k$ additional parameters needed to specify $\bV$ when one increases the dimension of $\bY$ by one from $k$. This collection of additional parameters can be viewed as the parameters needed to specify a $(k-1)$-sphere under a spherical coordinate system \citep{moon1988spherical}, in terms of both parameter counts and parameter interpretations. A spherical coordinate system for characterizing  $(k-1)$-spheres of arbitrary radius consists of one radial coordinate ranging over $[0, \, +\infty)$, where a radial parameter falls, one angular coordinate ranging over $[-\pi, \, \pi)$, which a longitude angle is within, and another $k-2$ angular coordinates, each ranging over $[0, \, \pi]$, which a latitude angle belongs to. These $k$ radial and orientation parameters can then link to $k$ parameters in $\mathbb{R}^k$ using the connection between the spherical coordinate system in the $(k-1)$-dimensional spherical space and the Cartesian coordinate system in the $k$-dimensional Euclidean space \citep{blumenson1960derivation}. This is the connection that relates  $\bgamma$ to $\widetilde\bOmega$ after partitioning $\widetilde\bOmega$ in a way that we demonstrate in a concrete example next. 

Suppose that $\bY\sim \mbox{ESAG}_4(\bmu, \, \bV)$ and thus $d=5$. After $\bmu$ is specified, we need  radial and orientation parameters in $\widetilde\bOmega=(r_1, r_2, r_3, \, \theta_1, \theta_2, \theta_3, \, \phi_1, \phi_2, \phi_3)^\T$ to specify $\bV$. We divide $\widetilde\bOmega$ into $3(=d-2)$ groups of parameters as follows:
\begin{itemize}
\item $(r_1, \, \theta_1)$, which are the only radial and orientation parameters needed to specify an ESAG resulting from normalizing a bivariate Gaussian random variable, i.e., $k=2$; 
\item $(r_2, \, \theta_2, \, \phi_1)$, which includes the three additional radial and orientation parameters as we move from a 2-dimensional ESAG to a 3-dimensional ESAG, i.e., the dimension of the random variable changes from $k=3$ to $k+1=4$; 
\item $(r_3, \, \theta_3, \, \phi_2, \, \phi_3)$, which contains the four additional radial and orientation parameters as the dimension of the ESAG random variable increases from $k=4$ to $k+1=5$.
\end{itemize}
In general, for $d\ge 3$, we divide $(d-2)(d+1)/2$ parameters in $\widetilde\bOmega$ into $d-2$ groups, with the first group being $(r_1, \,\theta_1)$, and, for $j=2,...,d-2$, the $j$-th group being $(r_j, \, \theta_j, \, \tilde \bphi_j)$, where  
$\tilde\bphi_j = (\tilde\phi_{j,1},...,\tilde\phi_{j,j-1})^\T$. In other words, $\tilde \phi_{j,k}$, for $j=2, \ldots, d-2$ and $k=1, \ldots, j-1$, are the original latitude angles assigned to the $j$-th group. We then formulate each group of parameters in $\widetilde\bOmega$ using a group of new parameters in the corresponding Euclidean space by invoking the connection between a spherical coordinate system and the corresponding Cartesian coordinate system \citep{blumenson1960derivation}. 

To adapt to the grouping for $\widetilde\bOmega$, we also define $\bgamma$ as $d-2$ groups of parameters, $\bgamma = (\tilde\bgamma_1^\T,...,\tilde\bgamma_{d-2}^\T)^\T$, where 
$\tilde\bgamma_j = (\gamma_{j,1},...,\gamma_{j,j+1})^\T\in \mathbb{R}^{j+1}$, for $ j=1,...,d-2$.
Then the transformations that map $\bgamma$ to $\widetilde\bOmega$ are given by $
r_1 = \|\tilde \bgamma_1\|, \, \, 
\theta_1 = \mbox{atan2}(\gamma_{1,2},\gamma_{1,1}), 
$ and, for $j=2, \ldots, d-2$, $r_j = \|\tilde\bgamma_{j}\|$, 
\begin{equation*}
\begin{aligned}
\theta_j & = \begin{cases}
0, & \text {if $\gamma_{j,j}^2+\gamma_{j,j+1}^2 = 0$,}\\
\displaystyle{\mbox{arccos}\frac{\gamma_{j,j}}{\sqrt{\gamma_{j,j}^2+\gamma_{j,j+1}^2}}},  & \text{if $\gamma_{j,j+1} \ge 0$ and $\gamma_{j,j}^2+\gamma_{j,j+1}^2 \neq 0$,}\\
\displaystyle{-\mbox{arccos}\frac{\gamma_{j,j}}{\sqrt{\gamma_{j,j}^2+\gamma_{j,j+1}^2}}},  & \text{if $\gamma_{i,i+1} < 0$}, \end{cases}\\
\tilde{\phi}_{j,k} & = \begin{cases}
0, & \text{if $\sum_{\ell=k}^{j+1} \gamma_{j,\ell}^2 = 0$, for $k=1, ..., j-1$ },\\
\displaystyle{\mbox{arccos} \frac{\gamma_{j,k} }{\sqrt{\sum_{\ell=k}^{j+1} \gamma_{j,\ell}^2 }}}, &  \text{otherwise, for $k=1, ..., j-1$.}
\end{cases}
\end{aligned}
\end{equation*}

This completes the parameterization of $\mbox{ESAG}_{d-1}(\bmu, \, \bV)$ so that all identifiable parameters in $\bOmega=(\bmu^\T, \, \bgamma^\T)^\T$  range over the entire real line. Having the parameter space being $\mathbb{R}^p$ greatly simplifies the implementation of maximum likelihood estimation for $\bOmega$. 

\section{Maximum likelihood estimation}
\label{sec:mle} 
Using the parameterization of ESAG developed in Section~\ref{sec:parameterization}, one can easily derive the likelihood function of a sample from ESAG, following which one can maximize the logarithm of it with respect to $\bOmega$ over $\mathbb{R}^p$ to obtain the maximum likelihood estimator (MLE) of $\bOmega$. Straightforward as it appears, some cautions should be given in this likelihood-based inference procedure, in part due to the nature of $\bgamma$. 

\subsection{Interpretations of parameters}
\label{sec:interp}
According to Section~\ref{sec:grouping}, $\bgamma=\bzero$ implies $\lambda_j=1$, for $j=1,...,d$, and thus $\bV=\bI_d$, leading to an isotropic hyperspherical distribution \citep{mardia2014statistics}. If $\bY \sim \mbox{ESAG}_{d-1}(\bmu,\, \bI_d)$, then, for any orthogonal matrix $\bP$ such that $\bP\bmu=\bmu$, we have $\bP\bY\sim \mbox{ESAG}_{d-1}(\bmu,\, \bI_d)$, i.e., $\bP\bY=\bY$ in distribution, or, $\bP\bY\stackrel{\mathcal{L}} {=}\bY$ in short. In addition, if $\tilde \bgamma_j=\bzero$, then $r_j=0$, and thus $\lambda_{j+1}=(r_j+1)\lambda_j=\lambda_j$, in which case we say that the distribution is isotropic in the subspace spanned by $\lbrace \bv_j,\bv_{j+1}\rbrace$, or partially isotropic. That is, given any orthogonal matrix $\bP$ such that $\bP\bmu =\bmu$ and $\bP\bv_k=\bv_k$, for $k\ne j, j+1$, we have $\bP\bY\stackrel{\mathcal{L}}{=}\bY$. Practically speaking, this means that rotating data from an isotropic (a partially isotropic) ESAG via certain orthogonal matrix that rotates the mean direction to itself (and rotates certain eigenvectors of $\bV$ to themselves) does not change the distribution of the data. From the modelling point of view, any level of isotropy of ESAG implies a reduced model. Hence, testing whether or not a data set can be modelled by a reduced, thus more parsimonious, ESAG amounts to testing hypotheses regarding  parameters in $\bgamma$. For example, testing $\bV=\bI_d$  is equivalent to testing $\bgamma=\bzero$. 

A note of caution one should bear in mind when obtaining the MLE of $\bOmega$ is that, 
 even though the mapping from  $\bgamma$ to  $(\lambda_2,\, \ldots, \, \lambda_{d-1}, \, \bv_1^\T, \, \ldots, \, \bv_{d-1}^\T)$ is bijective, the mapping from the latter to the former  is not a bijection because, as one can see in (\ref{eq:Vdecom}), if $\bv_j$ is an eigenvector of $\bV$ corresponding to the eigenvalue $\lambda_j$, then so is $-\bv_j$.  This suggests that there exist $\bgamma\ne \bgamma'$ yet both $\bgamma$ and $\bgamma'$ map to the same $\bV$ given $\bmu$. When this happens, we say that $\bgamma$ and $\bgamma'$ are equivalent. We show in Appendix A that, if $\bgamma$ and $\bgamma'$ are equivalent, then $\|\tilde\bgamma_j\| = \|\tilde\bgamma_j'\|$, for $j=1,...,d-2$, which in turn suggests that the interpretations of $\bgamma$ and $\bgamma'$ relevant to isotropy of ESAG are the same.
 
 A theoretical implication of the existence of equivalent $\bgamma$ and $\bgamma'$ is that, although one cannot claim consistency of the MLE of $\bgamma$ (since the MLE may consistently estimate $\bgamma$ or $\bgamma'$), the consistency of the MLE of $\bV$ is guaranteed by the invariance property of MLE \citep[Theorem 7.2.10,][]{casella2021statistical}. A numerical implication of this is that maximum likelihood estimation of $\bOmega$ tends to be very forgiving in terms of the starting value for $\bOmega$, especially when the focal point of inference lies in $\bmu$ and $\bV$. We provide empirical evidence of these implications in a simulation experiment next. 
 
\subsection{Empirical evidence}
\label{eq:mlesim}
Using the proposed parameterization, we generate a random sample of size $n=1000$ from $\mbox{ESAG}_3(\bmu, \bV)$, where $\bmu=(2,\,-2,\,-1,\,-3)^\T$, and $\bV$ is determined via $\bmu$ and $\bgamma=(\gamma_{1,1}, \, \gamma_{1,2}, \, \gamma_{2,1}, \, \gamma_{2,2}, \, \gamma_{2,3})^\T=(-2,\,5,\, 3, \,5, \,-8)^\T$.
We then maximize the log-likelihood function of this random sample to find the MLE of $\bOmega$, denoted by $\hat \bOmega$, using two different starting values of $\bOmega$: one coincides with the truth, the other is given by  $\bmu_0 = \bone_4$ and $\bgamma_0 = \bzero$. This produces two estimates of $\bOmega$. We repeat this experiment 100 times. In all 100 Monte Carlo replicates, we employ the Broyden-Fletcher-Goldfarb-Shanno  algorithm \citep{fletcher2013practical} to find a maximizer of the log-likelihood function. In fact, we find that most commonly used optimization algorithms work well in maximizing the objective function despite the choice of starting values, partly thanks to the fact that transformations involved in the parameterization derivations in Section~\ref{sec:parameterization} are mostly smooth and simple enough.
 
\begin{figure}[ht]
\begin{center}
\includegraphics[width=4.8in]{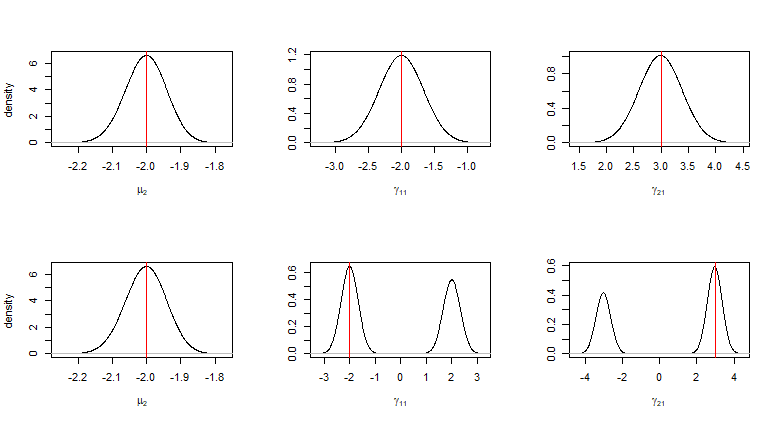}
\end{center}
\caption{\label{fig:mle}Estimated distributions of estimators for selected parameters in $\bOmega$ based on 100 realizations of each parameter estimator when the true parameter values are used as the starting value (upper panels) and when $\bmu_0$ and $\bgamma_0$ not equal to the truth are used as starting values (lower panels) in search for a maximizer of the log-likelihood. Vertical lines mark the true values of the corresponding parameters.}
\end{figure}

Figure~\ref{fig:mle} presents graphical summaries of 100  realizations of a subset of $\hat\bOmega=(\hat \bmu^\T, \hat \bgamma^\T)^\T$, $(\hat \mu_2, \, \hat \gamma_{1,1}, \, \hat \gamma_{2,1})$, corresponding to each choice of starting value. In particular, for each parameter, a kernel density estimate based on 100 realizations of its MLE is depicted in Figure~\ref{fig:mle}. The top panels of Figure~\ref{fig:mle}, which present results from using the truth of $\bOmega$ to start the optimization algorithm, provide empirical evidence suggesting that the usual asymptotic properties of an MLE, including consistency and asymptotic normality, are expected to hold for $\hat\bOmega$ when one uses a starting value in a neighborhood of the truth. The bottom panels of Figure~\ref{fig:mle}, which show results from using a starting value that has little resemblance with the truth, indicate that $\hat \bmu$ still behaves like a regular MLE that is consistent and asymptotically normally distributed, but $\hat \bgamma$ appears to follow a bimodal distribution. The two modes of the distribution of $\hat \bgamma$ are expected to be the true value of $\bgamma$ and another value $\bgamma'$ that is equivalent to $\bgamma$. 

\begin{figure}[ht]
\begin{center}
\includegraphics[width=4in]{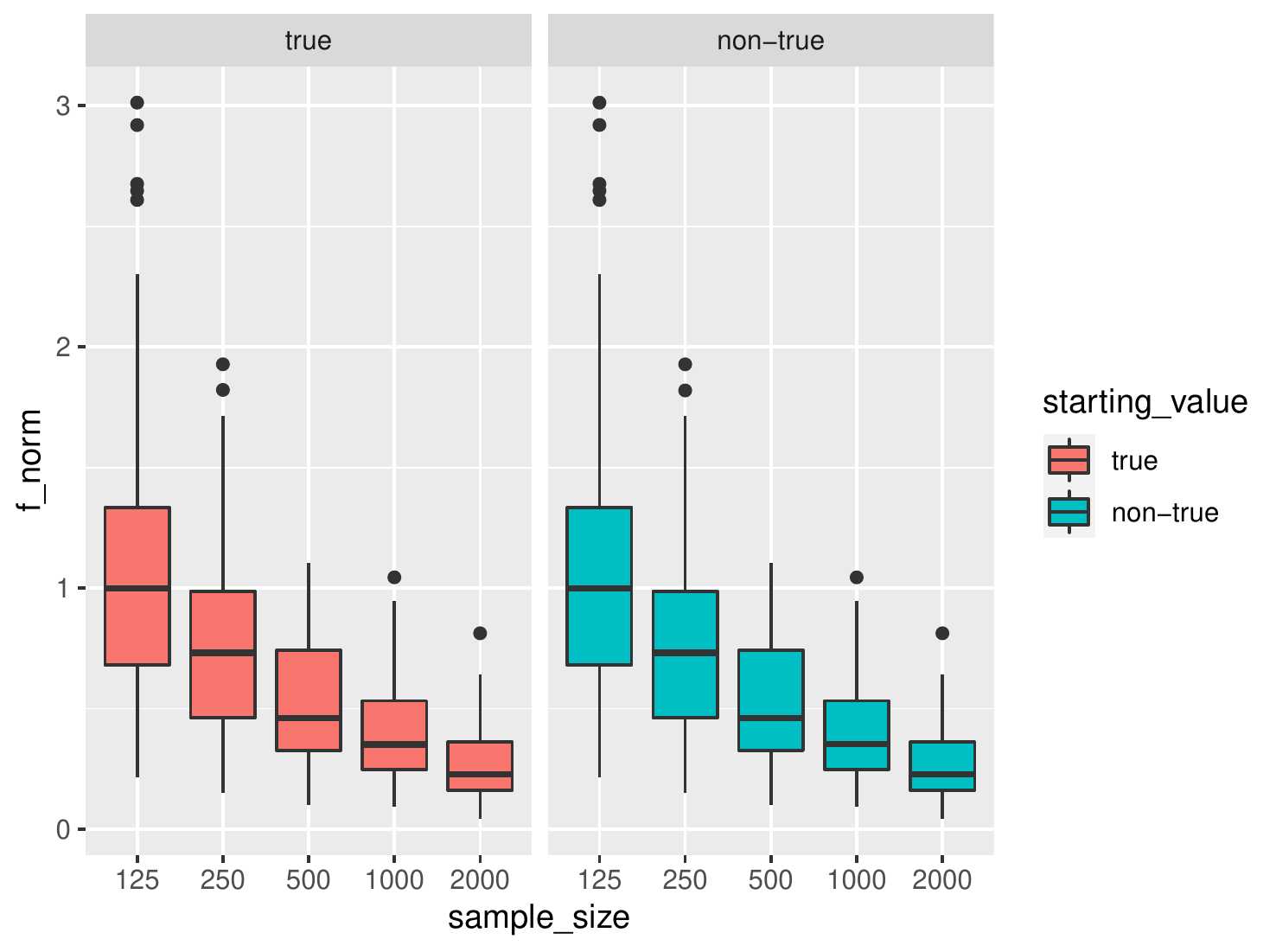}
\end{center}
\caption{\label{fig:V_consistency}Boxplots of the Frobenius norm of $\bV-\hat \bV$ as sample size $n$ varies when the true parameter values are used as the starting value (in the left panel) and when $\bmu_0$ and $\bgamma_0$ not equal to the truth are used as starting values (in the right panel) in search for a maximizer of the log-likelihood.}
\end{figure}

Despite the potential bimodality of $\hat \bgamma$ when a less carefully chosen starting value of $\bOmega$ is used to find $\hat \bOmega$, the resultant estimate of $\bV$, $\hat \bV$, is similar, if not identical, to the estimate one obtains when using the truth as the starting value. Figure~\ref{fig:V_consistency} shows boxplots of the Frobenius norm of $\bV-\hat \bV$ corresponding to 100 realizations of $\hat \bV$ resulting from each choice of the starting value. From there one can see that $\hat \bV$ is virtually unaffected by the choice of starting values. Although the robustness of $\hat \bmu$ and $\hat \bV$ to the choice of starting value is reassuring, one should not treat $\hat \bgamma$ as a conventional MLE due to its behavior observed in Figure~\ref{fig:mle}. Consequently, the usual Fisher information matrix or the sandwich variance does not serve well for estimating the variance of $\hat\bOmega$. We thus recommend use of  bootstrap for the uncertainty assessment of $\hat \bOmega$.

\section{Model diagnostics}
\label{sec:verify}
Even though the ESAG family accommodates certain anisotropic feature of a distribution and thus offers some flexibility in modelling, it remains fully parametric and thus is subject to model misspecification  in a given application. In this section, we develop residual-based model diagnostics tools that data analysts can use to assess whether or not an ESAG distribution provides adequate fit for their directional data, either as a marginal distribution, or a conditional distribution of the directional response given covariates $\bW$ as in a regression setting. 

\subsection{Residuals}
\label{sec:resid}
Denote by $\{\bY_i\}_{i=1}^n$ the observed directional data of size $n$, where $\bY_1, \ldots, \bY_n$ are independent with $\bY_i\sim \mbox{ESAG}_{d-1}(\bmu_i, \, \bV_i)$, for $i=1, \ldots, n$. The subscript $i$ attached to the mean and variance-covariance can be dropped if one aims to assess the goodness of fit (GOF) for the observed data using an ESAG as the marginal distribution. Otherwise the subscript implies covariate-dependent model parameters in ESAG as in a regression model for $\bY$. 

In a non-regression or regression setting, after one obtains the MLE of all unknown parameters in the model, one has the MLEs $\hat \bmu_i$ and $\hat \bV_i$, following which a prediction can be made by $\hat{\bY}_i  = \hat{\bmu}_i/\|\hat{\bmu}_i\|$, for $i=1, \ldots, n$. Similar to a directional residual defined in \citet{jupp1988residuals},  we define residuals as
\begin{equation}
\hat \br_i = \left(\bI_d-\hat \bY_i \hat \bY_i^\T\right)\bY_i, \mbox{ for $i=1, \ldots, n$}.\label{eq:ourresid}
\end{equation}
In (\ref{eq:ourresid}), $\hat{\bY}_i\hat{\bY}^\T_i$ can be viewed as the projection onto the space spanned by $\hat{\bmu}_i$, and thus $\bI_d-\hat \bY_i \hat \bY_i^\T$ is the projection onto the space orthogonal to the space spanned by $\hat{\bmu}_i$. Equivalently, by the orthogonality of eigenvectors of $\hat \bV_i$, $\bI_d-\hat \bY_i \hat \bY_i^\T$ is the projection onto the space spanned by the $d-1$ eigenvectors of $\hat \bV_i$ that are orthogonal to $\hat{\bmu}_i$, denote by $\lbrace \hat{\bv}_{i,j} \rbrace_{j=1}^{d-1}$. Hence (\ref{eq:ourresid}) can be re-expressed as $\hat \br_i=\hat \bP_{-d} \hat \bP_{-d}^\T \bY_i$, where  $\hat{\bP}_{-d}=[\hat{\bv}_{i,1} \mid ...\mid \hat{\bv}_{i, d-1}]$, that is, $\hat{\bP}_{-d}$ is the $d\times (d-1)$ matrix with the $j$-th column being $\hat{\bv}_{i,j}$, for $j=1,\ldots, d-1$. The potential dependence $\hat{\bP}_{-d}$ on covariates via the subscript $i$ is suppressed for simplicity. 

For model diagnostic purposes, we use the following quadratic form of residuals, 
\begin{equation}
\hat Q_i = \hat{\br}_i^\T \hat{\bV}_i^{-1} \hat{\br}_i, \mbox{ for $i=1, \ldots, n$}.
\label{eq:sqr}
\end{equation}
Note that $\hat \br_i=\hat \bP_{-d} \hat \bP_{-d}^\T \bY_i$  converges to $\br_i=\bP_{-d} \bP_{-d}^\T \bY_i$ in distribution, where $\bP_{-d}$ results from excluding the $d$-th column of the $d \times d$ matrix $\bP=[\bv_1 \mid ... \mid \bv_{d-1} \mid \bv_{d}]$, and $\bP_{-d}\bP_{-d}^\T = \bI_d - \bmu_i\bmu_i^\T/\|\bmu_i\|^2$. Additionally, $\hat{\bV}_i$ converges to $\bV_i$ in probability as $n\to \infty$. Thus, (\ref{eq:sqr}) converges to $ Q_i=\br_i^\T \bV^{-1}_i \br_i$ in distribution as $n\to \infty$. In what follows, we investigate the distribution of $Q_i$ to gain insight on the asymptotic distribution of (\ref{eq:sqr}). The subscript $i$ as the data point index is suppressed in this investigation. 

For $\bY \sim \mbox{ESAG}_{d-1}(\bmu, \, \bV)$, the random variable can be expressed as $\bY = \bX/\|\bX\| = (\bV^{1/2}\bZ+\bmu)/\|\bX\|$, where $\bZ \sim N_d(\bzero, \, \bI_d)$. Hence, $
\br = \bP_{-d}\bP_{-d}^\T\bY = {\bP_{-d}} \bP_{-d}^\T\bV^{1/2}\bZ/\|\bX\|$, following which
we show in Appendix B that 
\begin{align}
Q = \br^\T \bV^{-1} \br =\frac{\|\bU_{-d}\|^2}{\|\bX\|^2}, \label{eq:Q2}
\end{align}
where $\bU_{-d}$ results from replacing the $d$-th entry of $\bU=\bP^\T\bZ$ with zero. Since $\bP$ is an orthogonal matrix, $\bU=\bP^\T\bZ \sim N_d(\bzero, \, \bI_d)$, and thus $\|\bU_{-d}\|^2 \sim \chi_{d-1}^2 $.
Now we see that $Q$ relates to the quotient of norms of Gaussian vectors, the distribution of which was studied in \citet{miller1964distributions}, following which one can derive the distribution of $Q$ analytically. One then can see that $Q$ is not a pivotal quantity and its distribution is not of a form familiar or easy enough for direct use for model diagnosis. We next construct a transformation of $Q$ aiming at attaining an approximate pivotal quantity for the purpose of model diagnostics. 

\subsection{Graphical model diagnostic}
\label{sec:graphdiag}
Diagnostics methods proposed by \citet{paine2020spherical} and \citet{scealy2019scaled} build upon the finding that, if $\bY=(Y_1, \ldots, Y_d)^\T \sim \mbox{ESAG}_{d-1}(\bmu, \, \bV)$, then $\|\bmu\|(Y_1, \ldots, Y_{d-1})^\T$ converges in distribution to $N_{d-1}(\bzero, \, \sum_{j=1}^{d-1} \lambda_j^{-1} \bv_j \bv_j^\T)$ as $\|\bmu\|\to \infty$ \citep{paine2018elliptically}. Following this finding, one also has that $T_0= \|\bmu\|^2 Q =(\|\bmu\|^2/\|\bX\|^2) \|\bU_{-d}\|^2$
converges in distribution to $\chi_{d-1}^2$ for ESAG, and thus is a pivot in limit as $\|\bmu\|\to \infty$ (instead of $n\to \infty$). One may thus assess adequacy of a posited ESAG model for a data set by checking if $\{\hat T_{0,i}\}_{i=1}^n=\{\|\hat \bmu_i\|^2\hat Q_i\}_{i=1}^n$  approximately come from $\chi^2_{d-1}$. As seen in Figure~\ref{fig:ESAGdata}, a larger $\|\bmu\|$ implies that the distribution has a higher concentration and thus less variability in data. This diagnostic strategy based on $T_0$ is thus intuitively well motivated  since, with $\|\bmu\|$ large, $\|\bmu\|^2/\|\bX\|^2$ is expected to be close to one, making $T_0$ close to $\|\bU_{-d}\|^2\sim \chi_{d-1}^2$. However, empirical evidence from our extensive simulation study suggest that a practically unreasonably large $\|\bmu\|$ is needed to make $\chi_{d-1}^2$ a reasonably good approximation of the distribution of $T_0$. Consequently, this strategy based on $T_0$ is of little practical value since data observed in most applications can rarely have low enough variability to make this approximation satisfactory. 

\begin{figure}[ht]
\begin{center}
\includegraphics[width=3in]{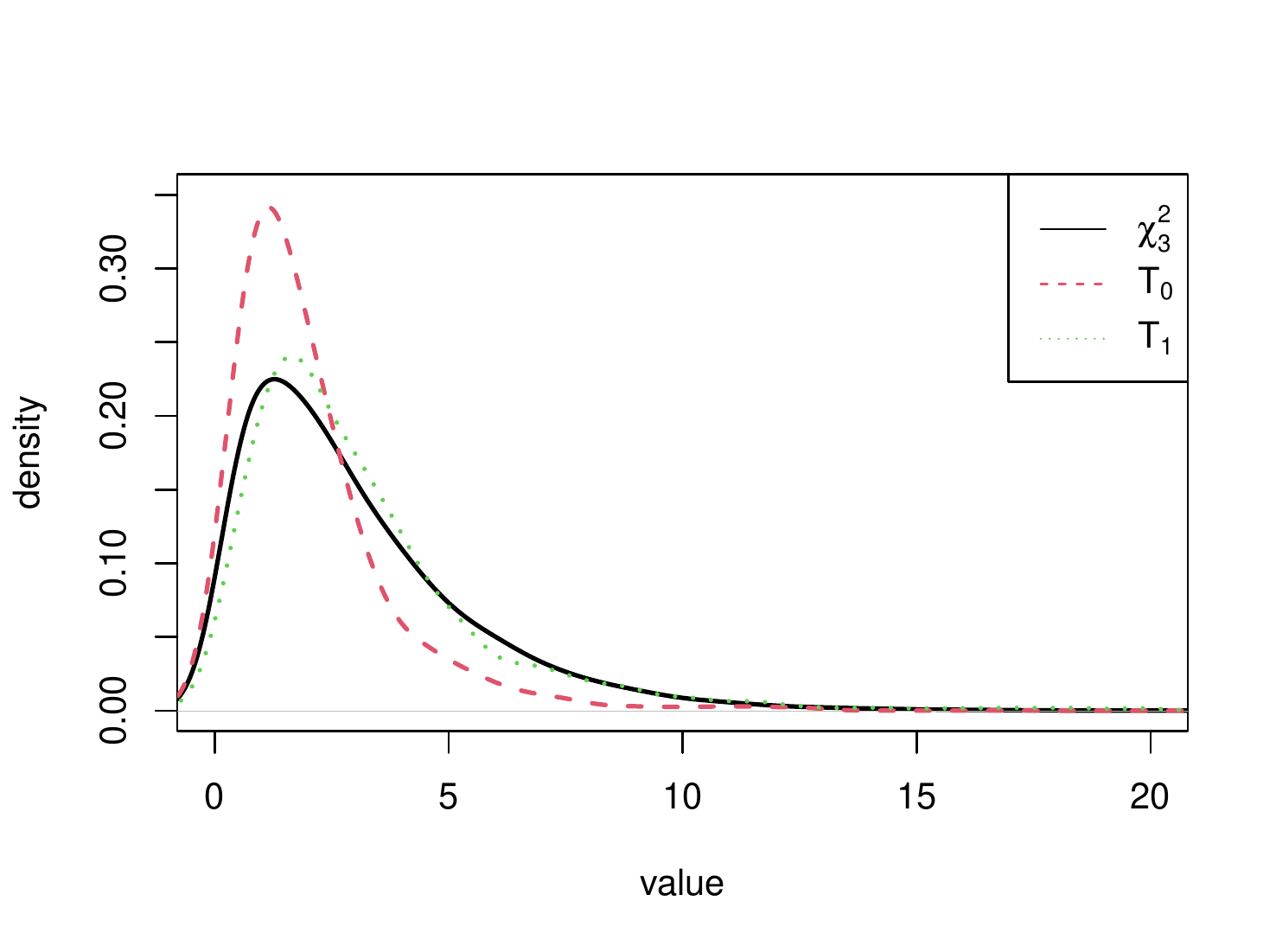}
\end{center}
\caption{Kernel density estimates of $T_0$ (dashed line) and $T_1$ (dotted line) comparing with the density of $\chi^2_3$ (solid line). \label{fig:T0T1T2}}
\end{figure}

Motivated by the fact that $E(\|\bX\|^2)=\|\bmu\|^2+\sum_{j=1}^d \lambda_j$ \citep[Theorem 5.2.1,][]{rencher2008linear}, we propose the following random quantity for diagnostics purposes, 
\begin{align}
T_1 & = \left(\|\bmu\|^2 + \sum_{j=1}^{d}\lambda_j\right) Q, \label{eq:newteststat2} 
\end{align}
which follows $\chi^2_{d-1}$ approximately when $\|\bmu\|$ is large, with the approximation improves much faster than that for $T_0$ as $\|\bmu\|$ increases, and thus is more like a pivot than $T_0$ is. Figure~\ref{fig:T0T1T2} presents kernel density estimates of the distributions of $T_0$ and $T_1$ based on random samples of these random quantities, each of size 500,  generated based on Monte Carlo replicates from $\mbox{ESAG}_3(\bmu, \, \bV)$. More specifically, we set $\|\bmu\|=4.24$, which is not large enough to make the $\chi^2$- approximation for $T_0$ satisfactory, and $\sum_{j=1}^{d}\lambda_j=11.1$. As one can see in this figure, the variability of $T_0$ is way too low to make $\chi^2_{d-1}$ approximate its distribution well, and $T_1$ greatly improves over $T_0$ in its proximity to $\chi^2_{d-1}$. In general, $T_1$ only requires a moderate $\|\bmu\|$ to make the $\chi^2$-approximation practically useful. 

Following maximum likelihood estimation of all unknown parameters, one can exploit an empirical version of $T_1$, $\{\hat T_{1,i}\}_{i=1}^n$, where $\hat T_{1,i}=(\|\hat\bmu_i\|^2 + \sum_{j=1}^{d}\hat\lambda_{i,j}) \hat Q_i$, for $i=1, \ldots, n$, and check if $\{\hat T_{1,i}\}_{i=1}^n$ can be reasonably well modeled by $\chi_{d-1}^2$. It can be a graphical check via a quantile-quantile (QQ) plot, for example, to see if there exists any clear signal of this sample deviating from $\chi_{d-1}^2$. Such graphical check is easy to implement following parameter estimation, and can provide visual warning signs when ESAG is a grossly inadequate model for the observed data $\{\bY_i\}_{i=1}^n$. Certainly, in a given application, the quality of $\chi^2$-approximation for $T_1$ is unknown with its true distribution yet to be estimated. We next propose a bootstrap procedure to facilitate a quantitative test for model misspecification, which leads to another graphical diagnostic tool as a byproduct that does not rely on a $\chi^2$-approximation for $T_1$. 

\subsection{Goodness of fit test}
\label{sec:GOF}
Consider testing the null hypothesis that $\bY$ follows an ESAG. Although $T_1$ defined in (\ref{eq:newteststat2})  approximately follows $\chi^2_{d-1}$ under the null hypothesis, a testing procedure based on $T_1$ that does not acknowledge its exact null distribution can lead to misleading conclusion, e.g., an inflated Type I error for the test. Instead of estimating the exact null distribution of $T_1$, we use a random sample of $T_1$ induced from an ESAG as a reference sample, and quantify the dissimilarity between this reference sample and the observed empirical version of $T_1$, $\{\hat T_{1,i}\}_{i=1}^n$. One may use a nonparametric test for testing if two data sets come from the same distribution, such as  the Kolmogorov–Smirnov (KS) test \citep{chakravarti1967handbook} and the Cram\'er-von Mises test \citep{anderson1962distribution}, to compare $\{\hat T_{1,i}\}_{i=1}^n$ and the reference sample induced from an ESAG. We employ the KS test in all presented simulation study in this article. A smaller $p$-value from the test indicates a larger distance between the underlying distribution of $\{\hat T_{1,i}\}_{i=1}^n$ and that of the reference sample, with the latter approximately representing what one expects for $T_1$ under the null hypothesis. Here, the ultimate test statistic for testing the null hypothesis is a $p$-value from the KS test. Denote this test statistic as $\mbox{KS}_p$. Even when data are from an ESAG, it is analytically unclear what $\mbox{KS}_p$ should be because the ESAG from which the reference sample is induced is not exactly the true ESAG (as to be seen next). We thus use parametric bootstrap to estimate the null distribution of $\mbox{KS}_p$ to obtain an approximate $p$-value to compare with a preset nominal level, such as 0.05, according to which we conclude to reject or fail to reject the null at the chosen nominal level. The following presents a detailed algorithm for this hypothesis testing procedure.

\begin{algorithm}[h!]\label{alg:gof}
\footnotesize
  \caption{\footnotesize  Goodness-of-Fit Test Procedure }\label{alg:gof test}
  \begin{algorithmic}[1]
    \Procedure{Compare observed empirical version of $T_1$ with a reference sample}{}
        \State Given data $\{\bY_i\}_{i=1}^n$ for a non-regression setting or $\lbrace(\bY_i,\bW_i)\rbrace_{i=1}^n$ for a regression setting, find the MLE $\hat{\bmu}_i$ and $\hat{\bgamma}_i$, for $i=1, \ldots, n$, assuming an ESAG model for $\bY_i$ or $\bY_i$ conditioning on $\bW_i$.
        \State Compute $\hat \bV_i$ and $\lbrace \hat{\lambda}_{i,j}\rbrace_{j=1}^{d-1}$ based on $\hat{\bmu}_i$ and $\hat{\bgamma}_i$, for $i=1, \ldots, n$.
        \State Compute $\hat T_{1,i}= (\|\hat{\bmu}_i\|^2+\sum_{j=1}^{d-1}\hat{\lambda}_{i,j})\hat Q_i$, for $i = 1 ,\ldots, n$. 
        \State Generate $\{\tilde\bY_i\}_{i=1}^n$, where $\tilde \bY_i \sim \mbox{ESAG}(\hat{\bmu}_i,\hat{\bV}_i)$, for $i = 1, ..., n$.
        \State Compute $\tilde T_{1,i}= (\|\hat{\bmu}_i\|^2+\sum_{j=1}^{d-1}\hat{\lambda}_{i,j})\tilde Q_i$, where $\tilde Q_i =\tilde r_i ^\T \hat \bV_i^{-1}\tilde r_i$ and $\tilde r_i=\hat \bP_{-d}\hat \bP_{-d}^\T \tilde \bY_i$, for $i = 1 ,... , n $.
        \State Use the KS test to test if $\lbrace \hat T_{1,i}\rbrace_{i=1}^n$ and $\lbrace \tilde T_{1,i}\rbrace_{i=1}^n$ arise from the same distribution. Denote by $\mbox{KS}_p$ the resultant $p$-value of the KS test. 
    \EndProcedure
    
    \Procedure{Bootstrap procedure to estimate the null distribution of $\mbox{KS}_p$}{}
        \State Set $B$ = number of bootstraps
        \State Initiate $s = 0$
        \For{$b$ in $1,...,B$}
        \State Generate the $b$-th bootstrap sample $\{\bY_i^{(b)}\}_{i=1}^n$, where $\bY_i^{(b)}\sim\mbox{ESAG}(\hat{\bmu}_i, \, \hat{\bV}_i)$ for $i = 1, ..., n$.
        \State Repeat steps 2--7 using data $\{\bY_i^{(b)}\}_{i=1}^n$ for a non-regression setting or $\lbrace(\bY_i^{(b)}, \bW_i)\rbrace_{i=1}^n$ for a regression setting. Denote the $p$-value of the KS test as $\mbox{KS}_p^{(b)}$. 
        \State \textbf{if} $\mbox{KS}_p^{(b)} < \mbox{KS}_p$ \textbf{then}  $s = s + 1$ 
      \EndFor
      \State  Define an estimated $p$-value for this GOF test as $s/B$.    \EndProcedure
  \end{algorithmic}
\end{algorithm}

Several remarks are in order for this algorithm. First, in Step 5, $\mbox{ESAG}(\hat \bmu_i, \, \hat\bV_i)$, from which we induce a data point $\tilde T_{1,i}$ in the reference sample $\{\tilde T_{1,i}\}_{i=1}^n$, can be viewed as the member of the ESAG family that is closest to the   distribution that characterizes the true data generating process producing $\bY_i$, where the closeness between two distributions is quantified by the Kullback-Leibler divergence \citep{white1982maximum}. Hence, $\tilde \bY_i$ generated from $\mbox{ESAG}(\hat \bmu_i, \, \hat\bV_i)$ at this step is expected to resemble $\bY_i$ if the null hypothesis is true, with $\hat \bmu_i$ and $\hat \bV_i$ consistently estimating $\bmu_i$ and $\bV_i$, respectively. Second, in Step 6, $\tilde T_{1,i}$ is constructed in a way that closely mimics $T_1$ instead of $\hat T_{1,i}$. In particular, just like $T_1$ where all population parameters are used in its construction, such as $\bmu$, $\{\lambda_j\}_{j=1}^d$, as well as $\bV$ and $\bP_{-d}$ that $Q$ depends on, computing $\tilde T_{1,i}$ (following steps 2--5) requires no parameter estimation although it depends on $\hat \bmu_i$, $\{\hat\lambda_{i,j}\}_{j=1}^d$, $\hat\bV_i$ and $\hat\bP_{-d}$, which are viewed as population parameters associated with $\tilde \bY_i$. One may certainly construct in Step 6 a random quantity closely mimicking $\hat T_{1,i}$ instead, but that would involve another round of parameters estimation based on $\{\tilde \bY_i\}_{i=1}^n$ and thus is  computationally unattractive. Third, we acknowledge that, even under the null hypothesis, $\mbox{ESAG}(\hat \bmu_i, \, \hat \bV_i)$ is not the true distribution of $\bY_i$, with MLEs in place of the true model parameters. Hence, even when the null hypothesis is true, $\{\hat T_{1,i} \}_{i=1}^n$ do not come from the same distribution as that of the reference sample $\{\tilde T_{1,i}\}_{i=1}^n$, but the two distributions are expected to be closer than when the null hypothesis is severely violated. The bootstrap procedure is designed to estimate the null distribution of the distance between these two distributions that is quantified by $\mbox{KS}_p$, with a smaller value of $\mbox{KS}_p$ indicating a larger distance and thus stronger  evidence against the null. As to be seen in the upcoming simulation study, this bootstrap procedure is capable of approximating the null distribution of $\mbox{KS}_p$ well enough to yield an empirical size of the test matching closely with any given nominal level.

In the absence of model misspecification, the distribution of $\{\tilde T_{1,i}\}_{i=1}^n$ approximates the distribution of $T_1$, with the accuracy of the approximation depends less on $\|\bmu\|$ than the $\chi^2$-approximation does. Therefore, a more reliable graphical diagnostic device than the aforementioned QQ plot using $\chi^2_{d-1}$ as a reference distribution is a QQ plot based on $\{\hat T_{1,i}\}_{i=1}^n$ and $\{\tilde T_{1,i}\}_{i=1}^n$, as we demonstrate in the upcoming empirical study.

\section{Simulation study}
\label{sec:sim}
\subsection{Design of simulation}
To demonstrate operating characteristics of the diagnostics methods proposed in Section~\ref{sec:verify}, we apply them to data $\{\bY_i\}_{i=1}^n$ generated according to four data generating processes specified as follows: 
\begin{itemize}
    \item[(M1)]An ESAG model, $\mbox{ESAG}_3(\bmu, \bV)$, with $\bmu=(2, \, -2, \, 3, \,-3)^\T$ and $\bV$ defined via $\bmu$ and $\bgamma=(2,\,3, \,5, \,8, \,2)^\T$.
    \item[(M2)]A mixture of ESAG and angular Cauchy, with a mixing proportion of $1-\alpha$ on $\mbox{ESAG}_3(\bmu, \bV)$ specified in (M1), where a random vector from an angular Cauchy is generated by normalizing a random vector from a multivariate Cauchy with mean $\bmu$. This creates a scenario where $(1-\alpha)\times 100\%$ of the data arise from EAG but the rest of the data deviate from ESAG, where $\alpha\in \{0.05, \, 0.1, \, 0.2\}$. 
    \item[(M3)]An angular Gaussian distribution,  $\mbox{AG}(\bmu, \, \tilde\bV)$, where $\det(\tilde \bV)=\alpha \ne 1$, which creates a scenario where the constraint in (\ref{eq:constraint2}) is violated. More specifically, when formulating (M1), one has the eigenvalues $\{\lambda_j\}_{j=1}^{d-1}$ and the corresponding eigenvectors $\{\bv_j\}_{j=1}^{d-1}$ of $\bV$, besides $\lambda_d=1$ and $\bv_d=\bmu/\|\bmu\|$. Using these quantities from (M1), we define $\tilde \bV=\sum_{j=1}^d \tilde\lambda_j\bv_j\bv_j^\T$, where $\tilde \lambda_j=\alpha^{1/(d-1)}\lambda_j$, for $j=1, \ldots, d-1$, and $\tilde \lambda_d=1$, with $\alpha\in \{0.05, 0.1, 5, 10\}$. Because $\tilde \bV\bmu=\bmu$, the constraint in (\ref{eq:constraint1}) for ESAG is satisfied for this angular Gaussian distribution.
    \item[(M4)]Similar to (M3) but $\tilde \lambda_j=\alpha^{-1/(d-1)}\lambda_j$, for $j=1, \ldots, d-1$, and $\tilde \lambda_d=\alpha\in \{0.1, 0.5, 2.5, 5\}$. This leads to $\tilde \bV \bmu=\alpha \bmu$ and thus violates constraint (\ref{eq:constraint1}). Because now $\mbox{det}(\tilde\bV)=1$, the constraint in (\ref{eq:constraint2}) for ESAG is satisfied for this angular Gaussian distribution.
\end{itemize}

We generate random samples of size $n\in \{250, 500, 1000\}$ following each data generating process. The proportions of data sets across 300 Monte Carlo replicates for which the GOF test rejects the null hypothesis at various significance levels are recorded for each simulation setting. This rejection rate estimates the size of the test under (M1), and sheds light on how sensitive the proposed diagnostic methods are to various forms and severity of deviations from ESAG exhibited in (M2)--(M4). We set $B=200$ in the bootstrap algorithm.

\subsection{Simulation results}
\begin{figure}[h]
\begin{center}
\includegraphics[width=3in]{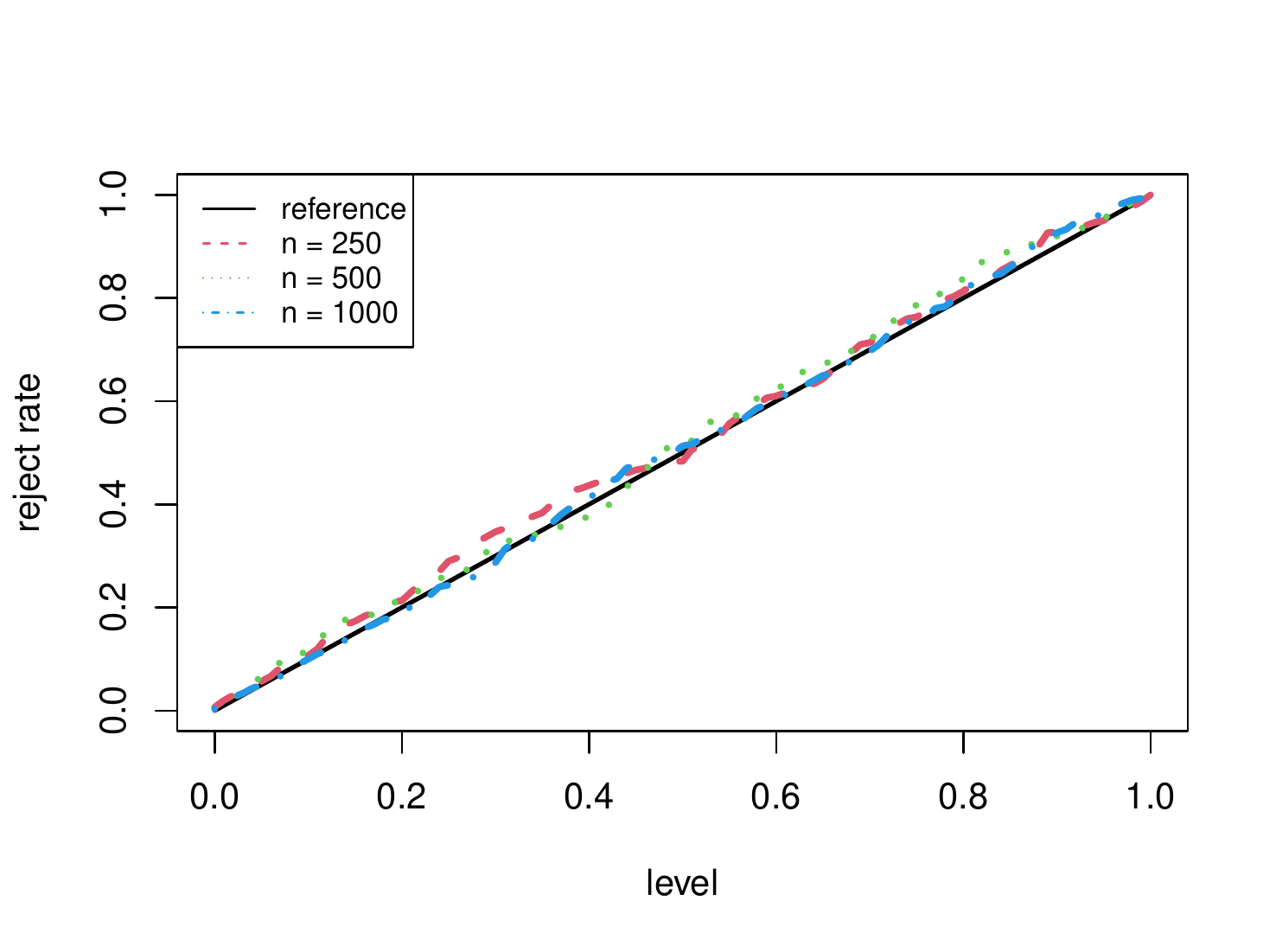}
\caption{\label{fig:M1}Rejection rates of the GOF test versus nominal levels under (M1) when $n=250$ (dashed line), 500 (dotted line), and 1000 (dash-dotted line). The solid line is the $45^\circ$ reference line.}
\end{center}
\end{figure}

Under (M1), Figure~\ref{fig:M1} shows the rejection rate versus the nominal level when the null hypothesis stating that $\bY\sim \mbox{ESAG}$ is true. This figure suggests that the null distribution of the test statistic $\mbox{KS}_p$ is approximated well enough over a wide range of nominal levels based on merely $B=200$ bootstrap samples, especially at the lower tail so that the size of the test is close to a low nominal level such as 0.05.

\begin{table}
\caption{\label{table:M2toM5}Rejection rates of the GOF test under (M2)--(M4) at nominal level 0.05}
\centering
\begin{tabular}{*{14}{c}}
\hline
$n$ & \multicolumn{3}{c}{(M2)} && \multicolumn{4}{c}{(M3)} && \multicolumn{4}{c}{(M4)} \\
\hline
$\{\alpha\}$ & 0.05 & 0.1  & 0.2 && 0.05 & 0.1 & 5 & 10 && 0.1 & 0.5 & 2.5 & 5\\ 
     \cline{2-4} \cline{6-9} \cline{11-14} 
250  & 0.10 & 0.27 & 0.65  && 0.27  & 0.17 & 0.14 & 0.17 && 0.47 & 0.16 & 0.75 & 1.00\\
500  & 0.17 & 0.42 & 0.89  && 0.38  & 0.30 & 0.22 & 0.33 && 0.73 & 0.26 & 0.98 & 1.00 \\
1000 & 0.26 & 0.69 & 0.99  && 0.60 & 0.46 & 0.30 & 0.52 && 0.96 & 0.42 & 1.00 & 1.00 \\
\hline
\end{tabular}
\end{table}
Table~\ref{table:M2toM5} presents rejection rates of the GOF test at nominal level 0.05 under the remaining three data generating processes (M2)--(M4). Under (M2), when $\alpha\times 100\%$ of the observed data are not from ESAG, the power of the test steadily increases as $\alpha$ increases. A larger sample size also boosts the power of detecting violation of the null. Under (M3), when data are from $\mbox{AG}(\bmu, \, \tilde \bV)$ that does not satisfy constraint (\ref{eq:constraint2}) due to $\det(\tilde \bV)=\alpha(\ne 1)$, one can see from Table~\ref{table:M2toM5} that, depending on the severity of the violation of (\ref{eq:constraint2}) that is controlled by the deviation of $\alpha$ from 1, the proposed test has a moderate power to detect this particular violation of ESAG, with a higher power at a larger sample size. Under (M4), when data are from $\mbox{AG}(\bmu,\tilde{\bV})$ with constraint (\ref{eq:constraint1}) violated due to $\tilde\bV\bmu=\alpha\bmu$, one can see from Table~\ref{table:M2toM5} that, as $\alpha$ deviates from 1 from either directions, the proposed test possesses moderate to high power to detect violation of the null hypothesis, with the power increasing quickly as $n$ grows larger. This can also serve as evidence for that, between the two constraints of ESAG in (\ref{eq:constraint1}) and (\ref{eq:constraint2}), violating the first constraint leads to an angular Gaussian deviating from ESAG more.

\begin{figure}[h]
\begin{center}
\includegraphics[width=2.3in]{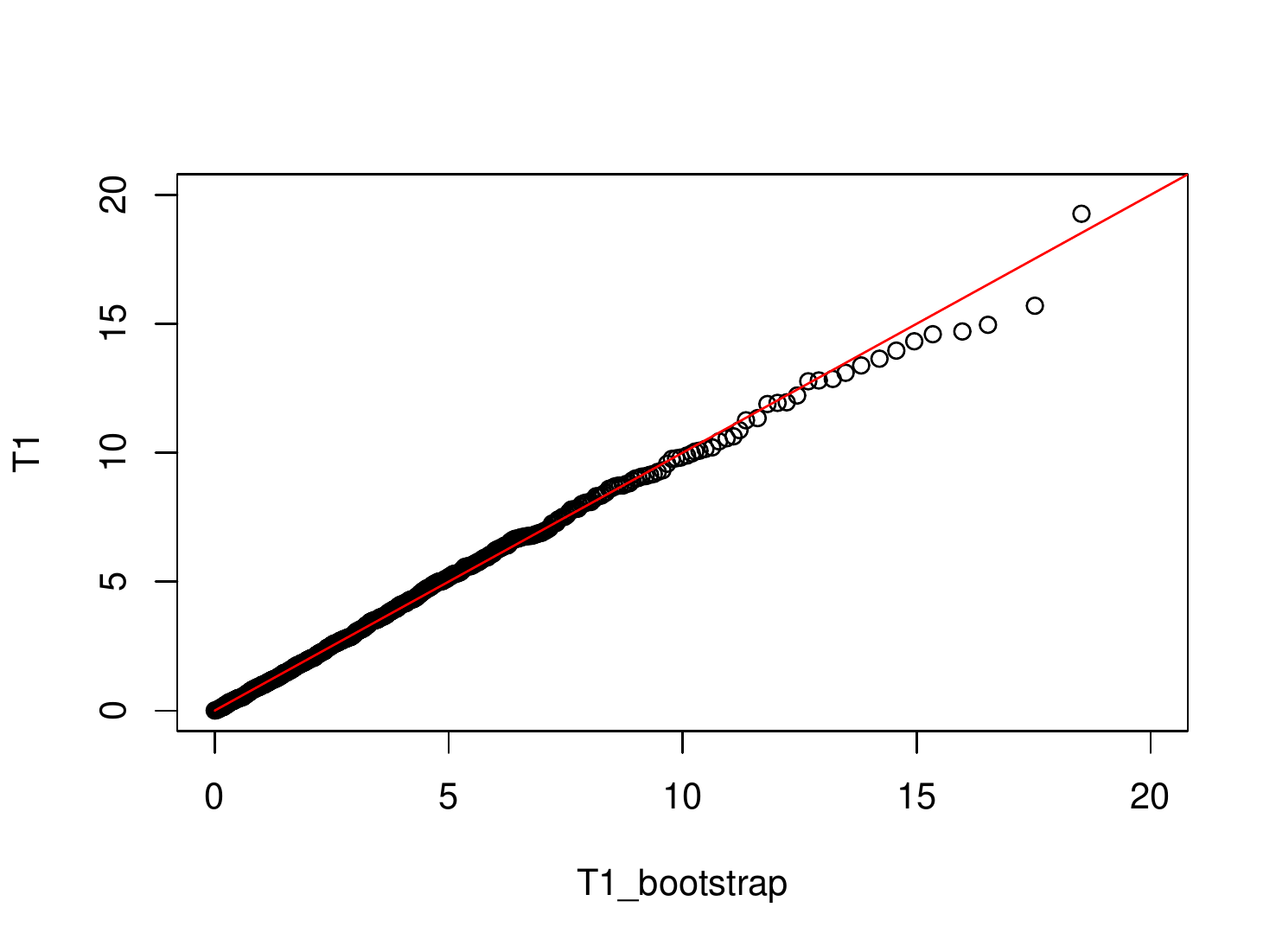}
\includegraphics[width=2.3in]{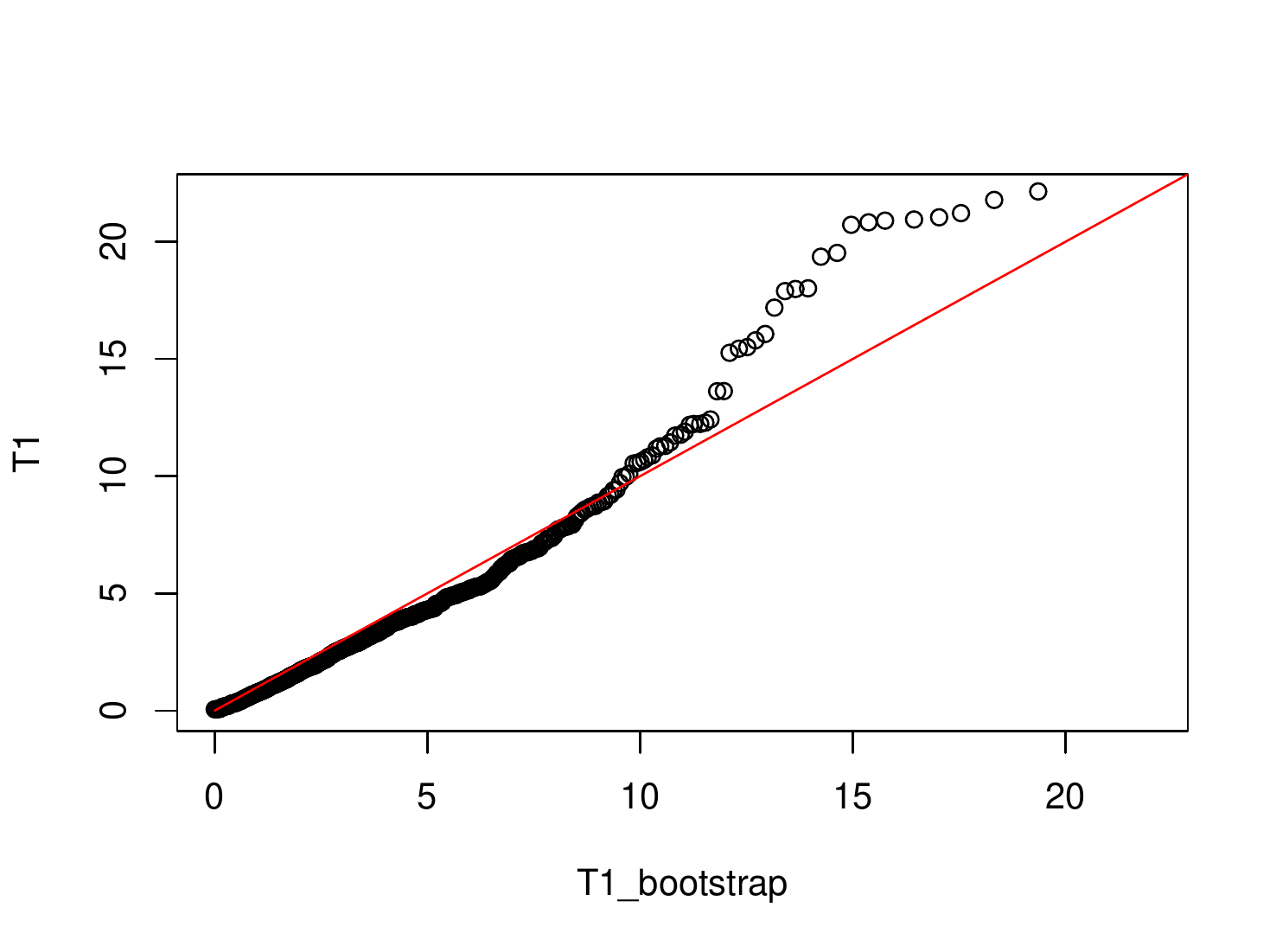}\\
\includegraphics[width=2.3in]{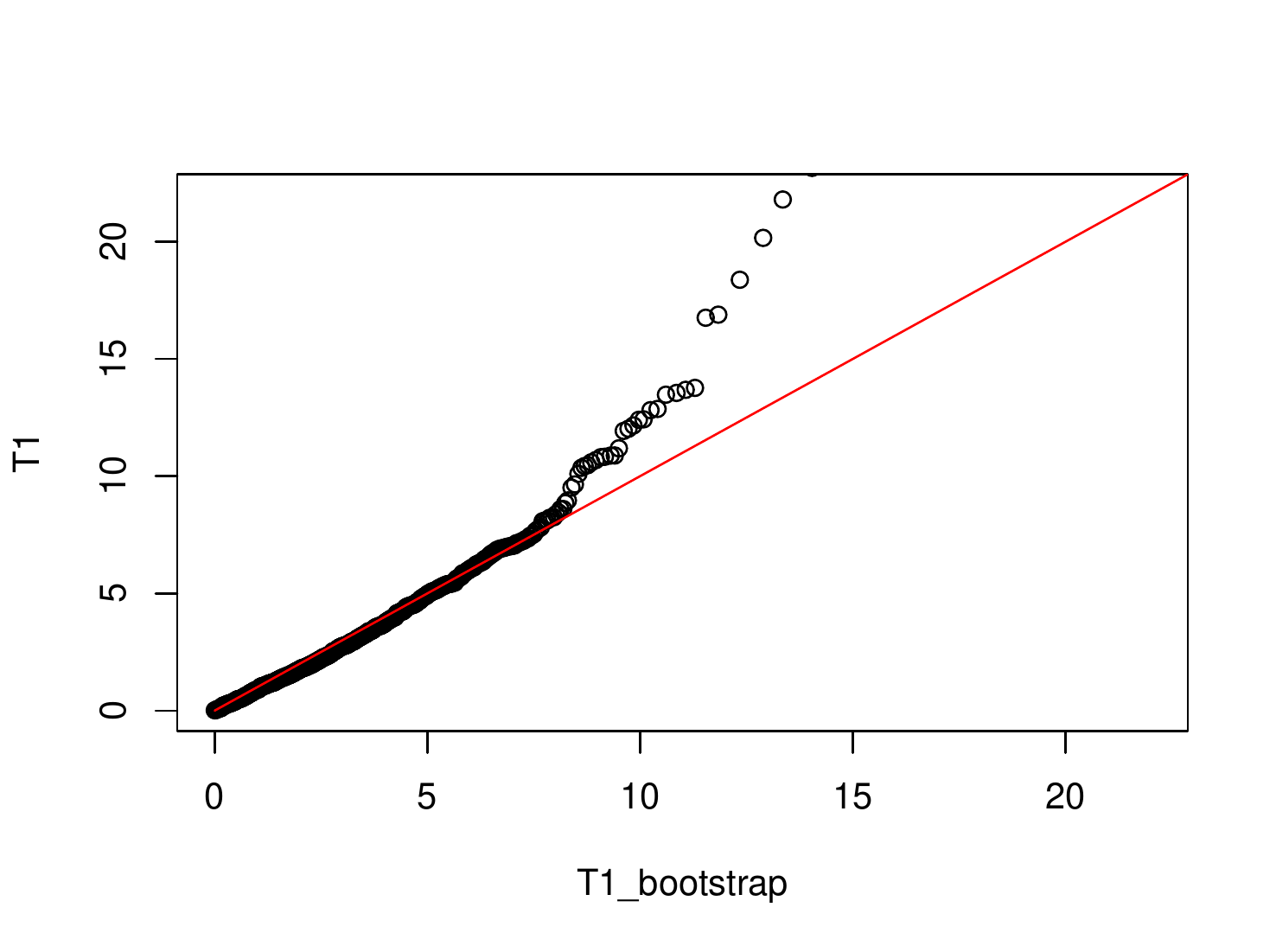}
\includegraphics[width=2.3in]{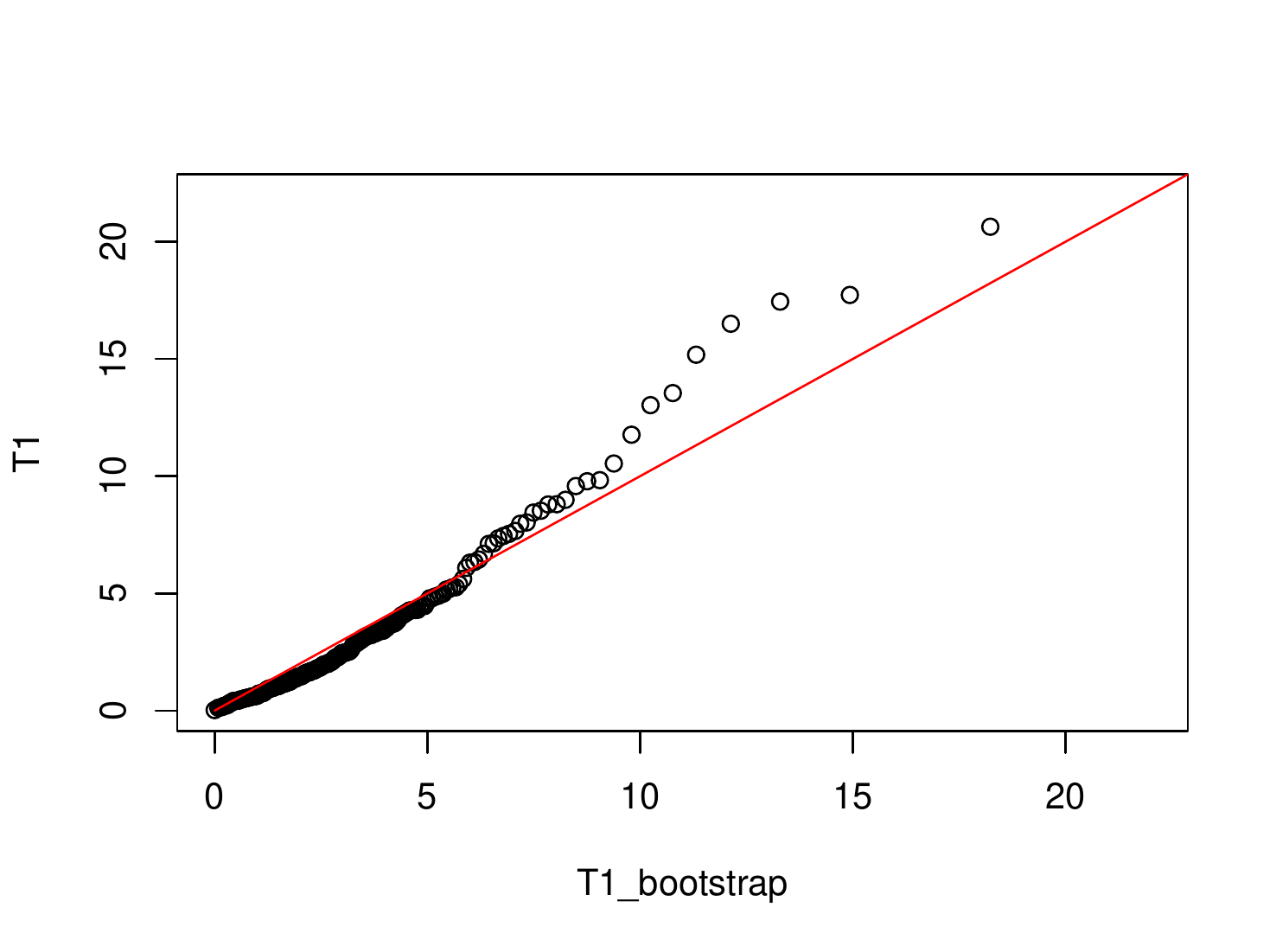}
\end{center}
\caption{ \label{fig:M1M2M3M4}QQ plots based on $\{\hat T_{1,i}\}_{i=1}^n$ and the bootstrap sample $\{\tilde T_{1,i}\}_{i=1}^n$ under (M1) (top-left panel), (M2) with $\alpha=0.2$ (top-right panel),  (M3) with $\alpha=0.05$ (bottom-left panel), and (M4) with $\alpha=2.5$ (bottom-right panel), respectively. Solid lines are $45^\circ$ reference lines.}
\end{figure}

Besides the quantitative GOF test that performs satisfactorily according to the above empirical evidence, one can also inspect the QQ plot based on $\{\hat T_{1,i}\}_{i=1}^n$ and the bootstrap sample $\{\tilde T_{1,i}\}_{i=1}^n$ to graphically check ESAG assumptions. Figure~\ref{fig:M1M2M3M4} shows a collection of such plots based on a randomly chosen Monte Carlo replicate from each of the four considered data generating processes. As evidenced in Figure~\ref{fig:M1M2M3M4}, violation of the ESAG assumptions as designed in (M2)--(M4) causes a QQ plot deviating from a straight-line pattern, a pattern more or less observed in the absence of model misspecification as in (M1). To create such QQ plots does not require the full $B$-round bootstrap procedure in the above algorithm, and provides a convenient graphical check on the  goodness of fit.

\section{Application to hydrochemical data}
\label{sec:realdata}
In this section, we analyze the hydrochemical data containing 14 molarities measured monthly at different stations along the Llobregat River and its tributaries in  northeastern Spain between the summer of 1997 and the spring of 1999  \citep{otero2005relative}. The complete data are available in the \texttt{R} package, \texttt{compositions} \citep{van2008compositions}. For illustration purposes, we focus on the compositional data recording relative abundance of two major ions, $\mbox{K}^+$ and $\mbox{Na}^+$, and two minor ions, $\mbox{Ca}^{2+}$ and $\mbox{Mg}^{2+}$. Taking the square-root transformation of the compostional data gives directional data with $d=4$. The four considered ions are mostly from potash mine tailing, which is one of the major sources of anthropogenic pollution in the Llobregat Basin \citep{soler2002sulfur}. 

\begin{figure}[h]
\begin{center}
\includegraphics[width=1.56in]{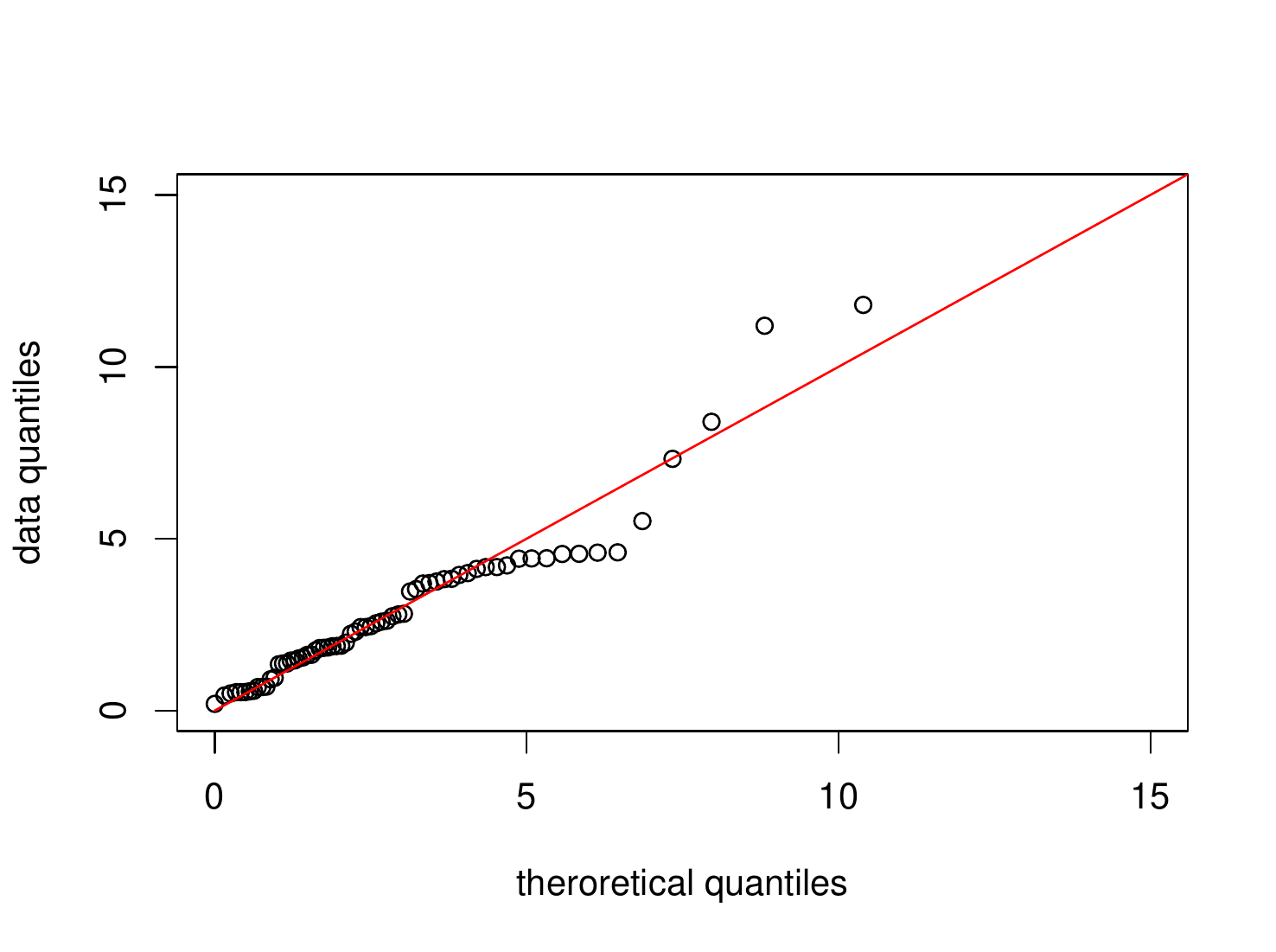}
\includegraphics[width=1.56in]{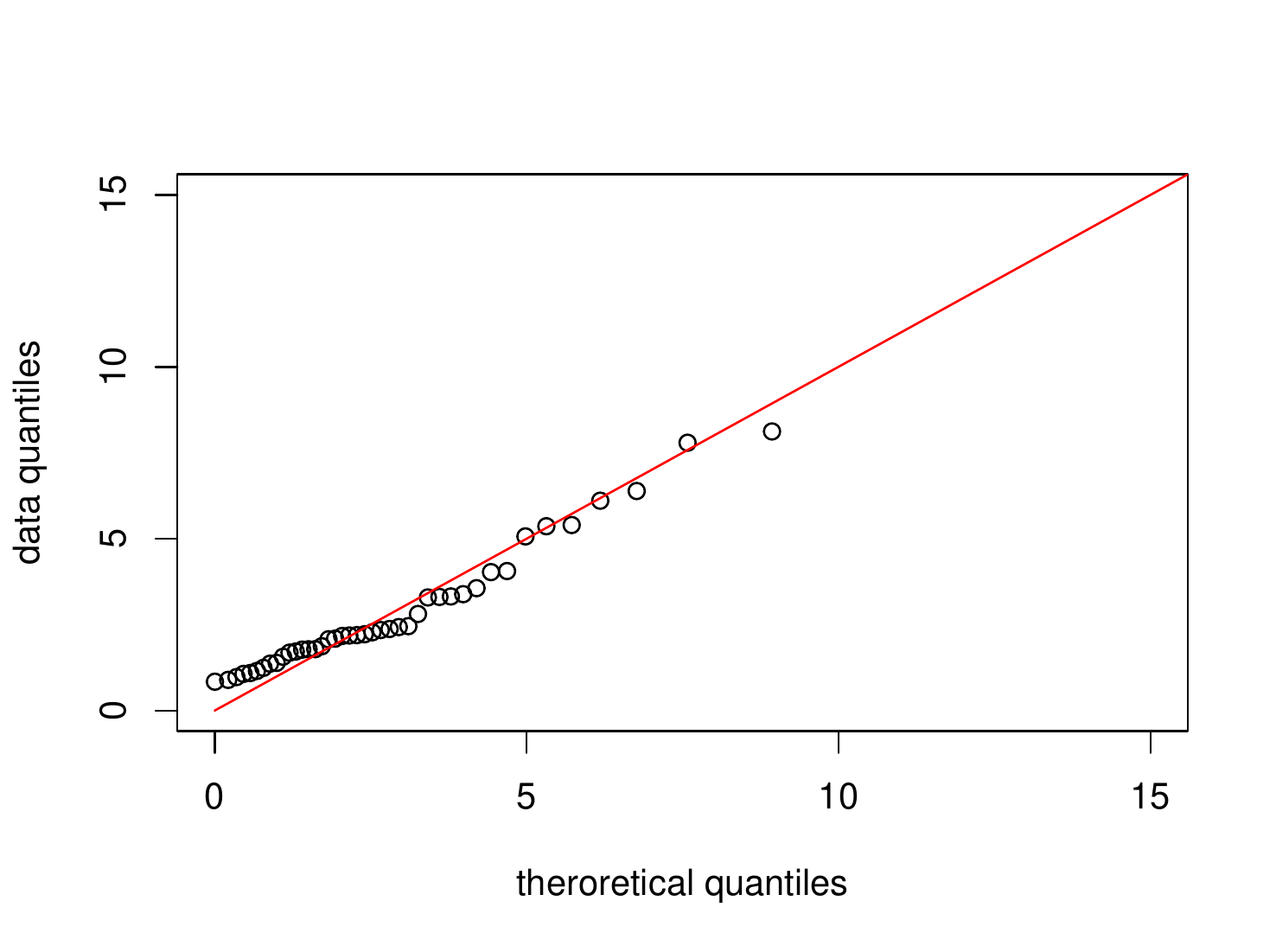}
\includegraphics[width=1.56in]{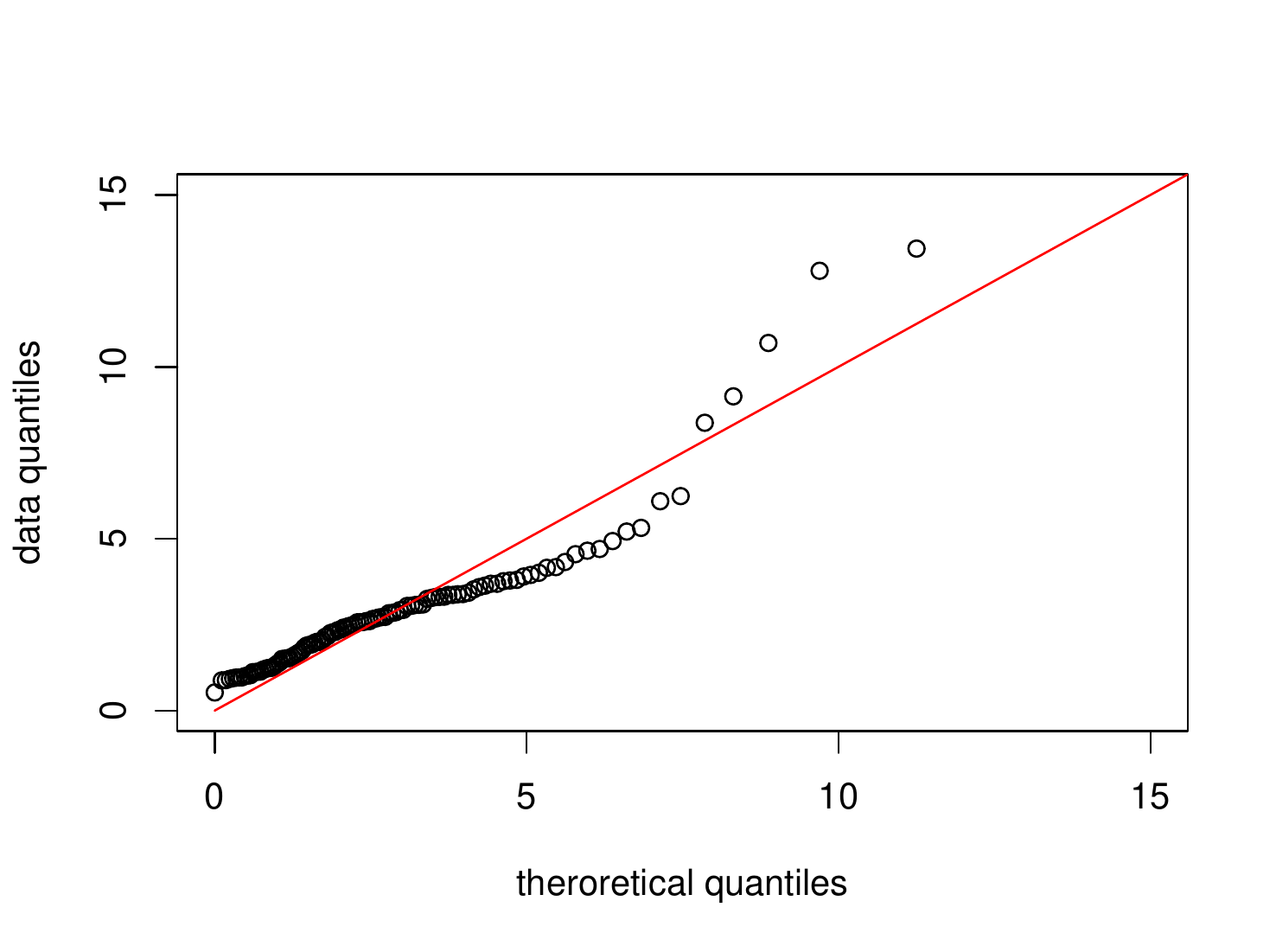}
\end{center}
\caption{QQ plots from the GOF test applied to compositional data from tributaries of Anoia (left panel), those from tributaries of the lower Llobregat course (middle panel), and the data that combine the previous two data sets (right panel). \label{fig:modelfit}}
\end{figure}

We first assume that the composition of $(\mbox{K}^+, \mbox{ Na}^+, \mbox{ Ca}^{2+},  \mbox{ Mg}^{2+})$ in tributaries of Anoia, one of the two main tributaries of the Llobregat River, follows an ESAG distribution. Using 67 records collected from stations placed along tributaries of Anoia, we obtain the estimated mean and variance-covariance of the compositional vector given by $$\hat{\bmu}_{\hbox {\tiny A}} = \begin{bmatrix}
1.99 \\
5.74  \\
7.95  \\
4.59  
\end{bmatrix}, 
\hspace{0.5cm} 
\hat{\bV}_{\hbox {\tiny A}}= \begin{bmatrix}
0.93 & 1.15 & -0.76 & -0.09 \\
1.15 & 2.77 & -1.41 & -0.27 \\
-0.76 & -1.41 & 1.99 & 0.38 \\
-0.09 & -0.27 & 0.38 & 0.73
\end{bmatrix}.$$ The GOF test yields an estimated $p$-value of 0.66, suggesting that the estimated ESAG distribution may provide an adequate fit for the data. The QQ plot in Figure~\ref{fig:modelfit} (see the left panel) may indicate some disagreement in the upper tail when it comes to the distribution of $\hat T_1$ and its bootstrap counterpart induced from an ESAG distribution, but otherwise mostly resemble each other in distribution. Transforming the estimated mean  $\hat{\bmu}_{\hbox {\tiny A}}$ back to the composition of four considered ions, we estimate the mean composition of $(\mbox{K}^+, \mbox{ Na}^+, \mbox{ Ca}^{2+},  \mbox{ Mg}^{2+})$ to be (0.03, 0.27, 0.52, 0.18). 

We repeat the above exercise for another compositional data of size 43 collected from stations placed along tributaries of the lower Llobregat course, and find the estimated mean vector and variance-covariance matrix to be 
$$\hat{\bmu}_{\hbox {\tiny L}} = \begin{bmatrix}
3.27 \\
8.56 \\
9.01 \\
5.78  
\end{bmatrix},
\hspace{0.5cm} 
\hat{\bV}_{\hbox {\tiny L}}= 
\begin{bmatrix}
0.63 & 1.50 & -0.71 & -0.90 \\
1.50 & 5.36 & -2.66 & -3.17 \\
-0.71 & -2.66 & 2.43 & 2.10 \\
-0.90 & -3.17 & 2.10 & 2.91
\end{bmatrix}.$$
The estimated $p$-value from the GOF test is 0.55 in this case. This, along with the QQ plot in Figure~\ref{fig:modelfit} (see the middle panel), also implies that the inferred ESAG distribution fits the data reasonably well. According to the estimated mean direction $\hat{\bmu}_{\hbox {\tiny L}}$, the estimated the mean composition of $(\mbox{K}^+, \mbox{ Na}^+, \mbox{ Ca}^{2+},  \mbox{ Mg}^{2+})$ is (0.05, 0.37, 0.41, 0.17), which shares some similarity with the estimated mean composition associated with Anoia tributaries in that $\mbox{Ca}^+$ and $\mbox{Na}^+$ are the two dominating components among the four, and $\mbox{K}^+$ is the minority. The two estimated variance-covariance matrices, $\hat{\bV}_{\hbox {\tiny A}}$ and $\hat{\bV}_{\hbox {\tiny L}}$, also share some implications in common: the two major ions, $\mbox{K}^+$ and $\mbox{Na}^+$, are positively correlated, so are the two minor ions, $\mbox{Ca}^{2+}$ and $\mbox{Mg}^{2+}$; but a major ion is negatively correlated with a minor ion in composition. 

Diagonal entries of $\hat{\bV}_{\hbox {\tiny A}}$ and $\hat{\bV}_{\hbox {\tiny L}}$ should not be interpreted or compared here in the same way as if data were not directional because the variability of ESAG($\bmu$, $\bV$) depends on both $\bmu$ and $\bV$. For the compositional vector as a whole, with $\|\hat \bmu_{\hbox {\tiny A}}\|\approx 11.00<\|\hat \bmu_{\hbox {\tiny L}}\|\approx 14.10$, we have data evidence suggesting that the compositional data from Anoia tributaries are less concentrated around its mean direction, and thus more variable, than those from tributaries of the lower Llobregat course. When zooming in on one component at a time in the compositional vector, one can compare variability between two ESAG distributions base on $\bV/\|\bmu\|^2$. For instance, even though $\hat \bV_{\hbox {\tiny A}}[3,3]=1.99<\hat \bV_{\hbox {\tiny L}}[3,3]=2.43$, we would not jump to the conclusion that the composition of $\mbox{Ca}^{2+}$ is less variable in Anoia tributaries than that in the other set of locations. Instead, because $\hat \bV_{\hbox {\tiny A}}[3,3]/\|\hat \bmu_{\hbox {\tiny A}}^2\|= 0.18>\hat\bV_{\hbox {\tiny L}}[3,3]/\|\hat \bmu_{\hbox {\tiny L}}^2\|= 0.17$, we conclude that the composition of $\mbox{Ca}^{2+}$ is similar in variability between the two sets of locations, but tributaries of Anoia may be subject to slightly higher variability in this regard. This  conclusion is also consistent with the comparison of the sample standard deviation of the composition of $\mbox{Ca}^{2+}$ between the two data sets. 

Moreover, estimates for the other set of parameters of ESAG arising in the new parameterization, $\bgamma$, also provide statistically interesting insights on the underlying distributions. Denote by $\hat\bgamma_{\hbox {\tiny A}}$ the estimate based on data from Anoia tributaries, and by $\hat\bgamma_{\hbox {\tiny L}}$ the estimate based on data from tributaries of the lower Llobregat course. We find that $\|\hat\bgamma_{\hbox {\tiny A}}\|=6.24 <\|\hat\bgamma_{\hbox {\tiny L}}\|=17.03$, indicating that neither of the two ESAG distributions is isotropic, with the second ESAG deviating from isotropy further. To check partial isotropy, we look into the estimated eigenvalues associated with $\hat \bV_{\hbox {\tiny A}}$ and $\hat \bV_{\hbox {\tiny L}}$. With one eigenvalue fixed at 1, the three estimated eigenvalues associated with $\hat \bV_{\hbox {\tiny A}}$ are 0.37 (0.05), 0.62 (0.10), and 4.44 (0.64), with the estimated standard errors in parentheses obtained based on 300 bootstrap data sets, each of the same size as the raw data sampled from the raw data with replacement. Similarly, we have the three estimated eigenvalues associated with $\hat \bV_{\hbox {\tiny L}}$ given by 0.19 (0.04), 0.54 (0.29), and 9.61 (1.84). Taking the estimated standard errors into consideration, with the large discrepancy between the estimated (and fixed) eigenvalues,  neither of the two data sets provides sufficient evidence indicating partial isotropy. 

Lastly, we fit the ESAG model to the 110 records that combine the above two data sets and obtain an estimated $p$-value of 0.02 from the GOF test, with the corresponding QQ plot clearly deviating from a straight line (see the right panel in Figure~\ref{fig:modelfit}). We thus conclude that an ESAG distribution is inadequate for modeling the data that mix compositional data from Anoia tributaries and those from tributaries of the lower Llobregat course. This lack of fit is not surprising because Anoia mostly passes through vineyards and industrialized zones,  whereas the Llobergat lower course also flows through densely populated areas with high demands of water besides agricultural and industrial areas. This explains the vastly different patterns and sources of anthorpogenic and geological pollution between Anoia and the lower Llobregat course \citep{gonzalez2012presence}, which create substantial heterogeneity in the mixed compositional data that an ESAG model is unlikely to capture. 

\section{Discussion}
\label{sec:conc}
Given the wide range of applications where directional data are of scientific interest and typically of dimension higher than three, an important first step towards sound statistical analysis of such data is the formulation of a directional distribution of arbitrary dimension. We adopt the initial formulation of the ESAG distribution proposed by  \citet{paine2018elliptically}, and take it to the next level via a sequence of reparameterizations leading to a distribution family indexed by parameters ranging over the entire real space. The resultant parametric family for directional data avoids pitfalls that many existing directional distributions suffer so that, unlike the Kent distribution for instance, there is no hard-to-compute normalization constant in the density function, and it is easy to simulate data from an ESAG of any dimension. More importantly, the proposed parameterization of ESAG lends itself to straightforward maximum likelihood inference procedures that are numerically stable and less dependent on ``good" starting values for parameter estimation. New parameters introduced along the way of reparameterization have statistically meaningful interpretations, which facilitate formulating hypothesis testing where one compares a reduced ESAG model, such as an isotropic or a partially isotropic model, with a saturated ESAG model. In summary, the proposed ESAG family of arbitrary dimension sets the stage for carrying out a full range of likelihood-based inference for directional data, including parameter estimation, uncertainty assessment, and hypothesis testing. 

To ease the concerns of model misspecification when assuming a parametric family in a given application, we develop graphical and quantitative diagnostics methods that utilize  directional residuals. Maximum likelihood estimation and the proposed diagnostics methods for ESAG can be easily implemented using the R code developed and maintained by the first author that is available upon request. 

An immediate follow-up step is to consider regression models for directional data, which is well motivated by the lack of fit of a marginal ESAG distribution for the mixed compositional data entertained in Section~\ref{sec:realdata}. We conjecture that, conditioning on covariates relating to geological features of considered tributaries and covariates reflecting human activities developed in regions these tributaries running through, the mixed compositioinal data can be better  modelled by an ESAG distribution with covariate-dependent $\bmu$ and $\bgamma$. With $\bmu$ and $\bgamma$ ranging over the entire real space of adequate dimensions, the proposed ESAG family prepares itself well for regression analysis of directional data without using complicated link functions to introduce dependence of model parameters on covariates $\bW$. For example, one may consider a fully parametric regression model as simple as $\bY|\bW\sim \mbox{ESAG}(\bmu(\bW), \, \bV(\bW))$, where $\bmu(\bW)$ is a linear function of covariates $\bW$, and $\bV(\bW)$ is determined by $\bmu(\bW)$ and $\bgamma(\bW)$, with the latter also a linear function of covariates. More flexible dependence structures of $\bmu$ and $\bgamma$ on covariates are also worthy of consideration in the follow-up research along the line of regression analysis. Once we enter the realm of regression models, the dimension of the parameter space grows more quickly as $d$ increases than before considering regression analysis for directional data. Upon completion of the study presented in this article, we have embarked on the exciting journey of developing scalable inference procedures suitable for settings with high dimensional parameter space following the strategies of frequentist penalized maximum likelihood estimation and Bayesian shrinkage estimation via hierarchical modeling.  


\par

\section*{Appendix A: Implication of $\bgamma$ and $\bgamma'$ being equivalent}
Under the proposed parameterization of $\mbox{ESAG}_{d-1}(\bmu, \bV)$, $\bV$ is determined by $\bgamma$ after $\bmu$ is specified. We thus write $\bV$ as $\bV(\bgamma)$ in this appendix, and view quantities related to $\bV$ as functions of $\bgamma$, such as the eigenvalues of $\bV$ and the radial parameters in (\ref{eq:lambdas}). If $\bgamma$ and $\bgamma'$ are equivalent, then  $\bV(\bgamma)=\bV(\bgamma')$, and thus $\bV(\bgamma)$ and $\bV(\bgamma')$ share the same eigenvalues. By (\ref{eq:lambdas}),  $\{\lambda_j(\bgamma)=\lambda_j(\bgamma')\}_{j=1}^{d-1}$ implies that $\{r_j(\bgamma)=r_j(\bgamma')\}_{j=1}^{d-2}$ . Lastly, from Section~\ref{sec:grouping},  $r_j=\|\tilde \bgamma_j\|$, for $j=1, \ldots, d-2$. Therefore, if $\bgamma$ and $\bgamma'$ are equivalent,  $\|\tilde \bgamma_j\|=r_j(\bgamma)=r_j(\bgamma')=\|\tilde \bgamma_j'\|$, for $j=1, \ldots, d-2$. 

\setcounter{equation}{0} 
\def\theequation{B.\arabic{equation}}
\section*{Appendix B: Proof of equation (\ref{eq:Q2})}
By the spectral decomposition theorem, $\bV^{\alpha}=\bP\bD^{\alpha}\bP^\T$, where $\bD^{\alpha} = \mbox{diag}(\lambda_1^\alpha,...,\lambda_d^\alpha)$ and $\bP=[\bv_1 \mid ... \mid \bv_{d}]$. Using this decomposition with $\alpha=-1$ and 1/2, we have 
\begin{align*}
Q & = \br^\T \bV^{-1} \br \\
& = \frac{\bZ^\T}{\|\bX\|} \bV^{1/2}\bP_{-d}\bP_{-d}^\T \times \bV^{-1} \times \bP_{-d}\bP_{-d}^\T \bV^{1/2}\frac{\bZ}{\|\bX\|} \\
& = \frac{\bZ^\T}{\|\bX\|^2} \bP\bD^{1/2} \bP^\T\bP_{-d}\bP_{-d}^\T \times \bP\bD^{-1} \bP^\T \times \bP_{-d}\bP_{-d}^\T \bP\bD^{1/2} \bP^\T \bZ, 
\end{align*}
where $$
\bP^\T\bP_{-d}  =
\begin{bmatrix}
\bP_{-d}^\T \\
\bv_d^\T
\end{bmatrix}
 \bP_{-d}=
 \begin{bmatrix}
 \bI_{d-1} \\ 
 \bzero^\T
 \end{bmatrix},
$$
and thus 
$$\bP^\T\bP_{-d} \bP_{-d}^\T \bP= 
\begin{bmatrix}
 \bI_{d-1} & \bzero \\ 
 \bzero^\T & 0
 \end{bmatrix}\triangleq \tilde \bI_d.$$
It follows that 
\begin{align*}
Q & =\frac{1}{\|\bX\|^2} \bZ^\T \bP \bD^{1/2}  \tilde \bI_d  \bD^{-1}  \tilde \bI_d \bD^{1/2} \bP^\T\bZ \\
& =
\frac{1}{\|\bX\|^2} \bZ^\T\bP   \tilde \bI_d \bD^{1/2}\bD^{-1}\bD^{1/2}   \tilde \bI_d \bP^\T \bZ\\
& = \frac{1}{\|\bX\|^2} \bU^\T  \tilde \bI_d \tilde \bI_d\bU, \mbox{ where $\bU=\bP^\T\bZ$,}\\
& = \frac{1}{\|\bX\|^2} \bU_{-d}^\T \bU_{-d}, \mbox{ where $\bU_{-d}= \tilde \bI_d  \bU$,}
\end{align*}
which gives (\ref{eq:Q2}).

\bibliographystyle{apalike}
\bibliography{esagref}
\end{document}